\documentclass[%
  aip,%
 amsmath,amssymb,floatfix,
reprint,%
twocolumn,%
jcp,%
]{revtex4-1} 

\usepackage{amsmath,amsthm,latexsym,amssymb,amsfonts,epsfig}

\usepackage[english]{babel}
\usepackage[utf8]{inputenc}			
\usepackage{graphicx, texdraw}  
\usepackage{color}
\usepackage{bm}

\usepackage{type1ec}
\usepackage{mathtools}
\usepackage{mathrsfs} 

\usepackage[margin=0.56in]{geometry}  

\usepackage{enumerate}

\usepackage[colorlinks=true,citecolor=OliveGreen, linkcolor=blue]{hyperref}

\usepackage{csquotes} 

\usepackage{color}
\usepackage[svgnames,dvipsnames]{xcolor}

\usepackage{amsmath}
\usepackage{amsfonts}
\usepackage{amssymb}
\usepackage{braket}

\newcommand{\vect}[1]{\boldsymbol{#1}}

\newcommand{\bNabla}{\boldsymbol{\nabla}}

\newcommand{\bPi}{\boldsymbol{\Pi}}

\newcommand{\kS}{\kappa_\mathrm{S}}
\newcommand{\kA}{\kappa_\mathrm{A}}
\newcommand{\kB}{\kappa_\mathrm{B}}

\newcommand{\kBolt}{k_\mathrm{B}}

\newcommand{\alphaB}{\alpha_\mathrm{B}}

\newcommand{\betaB}{\beta_\mathrm{B}}

\newcommand{\TS}{T_\mathrm{S}}

\newcommand{\vp}{\varphi}

\newcommand{\G}{\mathcal{G}}
\newcommand{\R}{\vect{r}}
\newcommand{\Intd}{\mathrm{d }}

\newcommand{\F}{\vect{F}}

\newcommand{\X}{\vect{x}}
\newcommand{\vecT}{\vect{t}}

\newcommand{\T}{\vect{t}}

\newcommand{\bv}{\vect{v}}

\newcommand{\dd}{\vect{d}}

\newcommand{\Nabla}{\boldsymbol{\nabla}}

\newcommand{\E}{\operatorname{E}}
\newcommand{\bigO}{\mathcal{O}}

\newcommand{\eR}{\vect{e}_r}
\newcommand{\ePhi}{\vect{e}_{\phi}}
\newcommand{\eThe}{\vect{e}_{\theta}}

\newcommand{\gOne}{\vect{g}_1}
\newcommand{\gTwo}{\vect{g}_2}

\newcommand{\Fext}{\vect{F}_2}

\newcommand{\JS}{J_{\mathrm{S}}}

\newcommand{\Gmatr}{\boldsymbol{\mathcal{G}}}

\newcommand{\vStok}{\vect{v}^{\mathrm{S}}}
\newcommand{\vStokcom}{{v}^{\mathrm{S}}}
\newcommand{\xOne}{\vect{x}_1}
\newcommand{\xTwo}{\vect{x}_2}

\newcommand{\vecD}{\vect{d}}
\newcommand{\infSum}{\sum_{n=0}^{\infty}}
\newcommand{\infSumOne}{\sum_{n=1}^{\infty}}

\newcommand{\zB}{z_{\mathrm{B}}}

\newcommand{\bgamma}{\boldsymbol{\gamma}}
\newcommand{\bGamma}{\boldsymbol{\Gamma}}
\newcommand{\bPsi}{\boldsymbol{\Psi}}

\newcommand{\EB}{E_\mathrm{B}}

\newcommand{\avg}[1]{\left\langle #1 \right\rangle}
\newcommand{\td}[2]{\frac{\mathrm{d} #1}{\mathrm{d} #2}} 

\definecolor{OliveGreen}{rgb}{0,0.6,0}
 \definecolor{darkGreen}{rgb}{0,0.4,0}
 \definecolor{darkBlue}{rgb}{0.,0.,0.6}

\begin{document}
\title{Hydrodynamic mobility of a solid particle nearby a spherical elastic membrane. II. Asymmetric motion}

\author{Abdallah Daddi-Moussa-Ider}
\affiliation
{Biofluid Simulation and Modeling, Fachbereich Physik, Universit\"at Bayreuth, Universit\"{a}tsstra{\ss}e 30, Bayreuth 95440, Germany}

\author{Maciej Lisicki}
\affiliation
{Department of Applied Mathematics and Theoretical Physics, Wilberforce Rd, Cambridge CB3 0WA, United Kingdom}
\affiliation
{Institute of Theoretical Physics, Faculty of Physics, University of Warsaw, Pasteura 5, 02-093 Warsaw, Poland }

\author{Stephan Gekle}
\affiliation
{Biofluid Simulation and Modeling, Fachbereich Physik, Universit\"at Bayreuth, Universit\"{a}tsstra{\ss}e 30, Bayreuth 95440, Germany}

\date{\today}

\begin{abstract}

In this paper, we derive analytical expressions for the leading-order hydrodynamic mobility of a small solid particle undergoing motion tangential to a nearby large spherical capsule whose membrane possesses resistance towards shearing and bending.
Together with the results obtained in the first part (Daddi-Moussa-Ider and Gekle, Phys. Rev. E {\bfseries 95}, 013108 (2017)) where the axisymmetric motion perpendicular to the capsule membrane is considered, the solution of the general mobility problem is thus determined.
We find that shearing resistance induces a low-frequency peak in the particle self-mobility, resulting from the membrane normal displacement in the same way, although less pronounced, to what has been observed for the axisymmetric motion.
In the zero frequency limit, the self-mobility correction near a hard sphere is recovered only if the membrane has a non-vanishing resistance towards shearing.
{We further compute the particle in-plane mean-square displacement of a nearby diffusing particle, finding that the membrane induces a long-lasting subdiffusive regime.}
Considering capsule motion, we find that the correction to the pair-mobility function is solely determined by membrane shearing properties.
Our analytical calculations are compared and validated with fully resolved boundary integral simulations where a very good agreement is obtained.

\end{abstract}
\maketitle

\section{Introduction}

Transport processes on the microscale play a key role in many biological and industrial applications \cite{wang12nano, gao12}. Typical examples include drug delivery involving nanoparticles required to reach specific areas of patients' bodies \cite{singh09, naahidi13}, problems of blood circulation \cite{tan12, gao14, hofmann15}, and also motion in crowded cellular environments \cite{zhou09, chen11, hoefling13}. A common feature of these processes is the presence of nearby interfaces, thus the motion occurs predominantly in geometric confinement. In living systems, the confining boundaries often possess a certain degree of elasticity which introduces additional memory effects to the system \cite{bickel06, bickel14, daddi16}. 

At small length scales, aqueous systems are typically characterized by a negligibly small Reynolds number, and the resulting overdamped motion can therefore be accurately described within the framework of linear Stokes equations \cite{happel12, kim13}. 
The relations between forces and velocities of particles in flow are therefore linear and quantified by the hydrodynamic mobility coefficients. They are determined by the long-range, fluid-mediated hydrodynamic interactions.

In this work, we focus on the case of a small colloidal particle translating under the action of a force in the presence of a nearby large spherical elastic capsule. This system may be looked upon as a simplistic model of transport of colloids close to cellular membranes \cite{kress05, chen13, juenger15}. Our aim is to assess the effects of elasticity on the motion of the particle itself, and also on the deformable capsule. 
A similar problem has been examined before by  Fuentes and coworkers \cite{fuentes88, fuentes89}, who have treated analytically the case of interactions between two unequal spherical drops at moderate separations. 
Being purely viscous, however, that system does not possess a memory and thus leads to hydrodynamic mobilities which are independent of frequency.
Their idea of solution relied on the image singularities technique, i.e. finding an appropriate system of images for a given distribution of forces outside a spherical droplet. 
Inspired by this work, we aim to find the analytical expression for the Green's function for a point-force near a spherical capsule. 
The surface of the capsule is made of an elastic membrane \cite{barthes16}, which resists against shearing and bending deformation, and is modeled using the combined Skalak \cite{skalak73} and Helfrich \cite{helfrich73} models. 
This model has been successfully used in our previous works for the case of confinement by one \cite{daddi16, daddi16c, daddi17} or two planar membranes \cite{daddi16b}.
Further theoretical investigations near elastic interfaces have been carried out via soft lubrication theory \cite{salez15, saintyves16, rallabandi16}.

In the preceding paper \cite{daddi17b} (hereafter referred to as part~I), we have derived the expression for the Green's function in the case when the point-force is directed along the symmetry axis, joining the centers of the particle and the capsule. In this contribution, we extend these results by providing a direct solution for the case when the point-force acts tangentially to the surface of the membrane, 
thus determining together the solution of the general mobility problem.
The Green's function is then used to evaluate the frequency-dependent self-mobility of a small particle moving close to the capsule, and the pair-mobility, which quantifies the effect of the force on the motion of the capsule itself. The solution is also used to compute the resulting deformation of the spherical capsule. The theoretical predictions at zero frequency are in agreement with the hard-sphere limit provided that the membrane possesses a non-vanishing resistance towards shearing.
Our analytical results comply with fully resolved boundary integral simulations which we have performed to validate the model.

The paper is organized as follows. 
In Sec.~\ref{sec:singularitySolution}, the solution of the fluid motion inside and outside the elastic capsule is expressed in terms of multipole expansions.
In Sec.~\ref{sec:particleSelfMobility}, analytical expressions of the particle frequency-dependent self-mobility nearby a membrane with pure shearing or pure bending are obtained in the point-particle framework and expressed in terms of infinite but convergent series.
{We compute in Sec.~\ref{sec:diffusion} the particle in-plane mean-square displacement, finding that the membrane introduces a memory in the system, leading at intermediate time scales of motion to a subdiffusive behavior of the nearby particle.}
Capsule motion and membrane deformation are computed in Sec.~\ref{sec:capsuleMotionAndDeformation}.
In Sec.~\ref{sec:comparisonWithBIM}, a comparison between analytical predictions and fully resolved boundary integral simulations is made where a very good agreement is obtained.
Concluding remarks are offered in Sec.~\ref{sec:conclusions}.
{Mathematical details which are not essential to understand our approach are given in the appendices.}

\begin{figure}
\begin{center}
\includegraphics[width=0.3\textwidth]{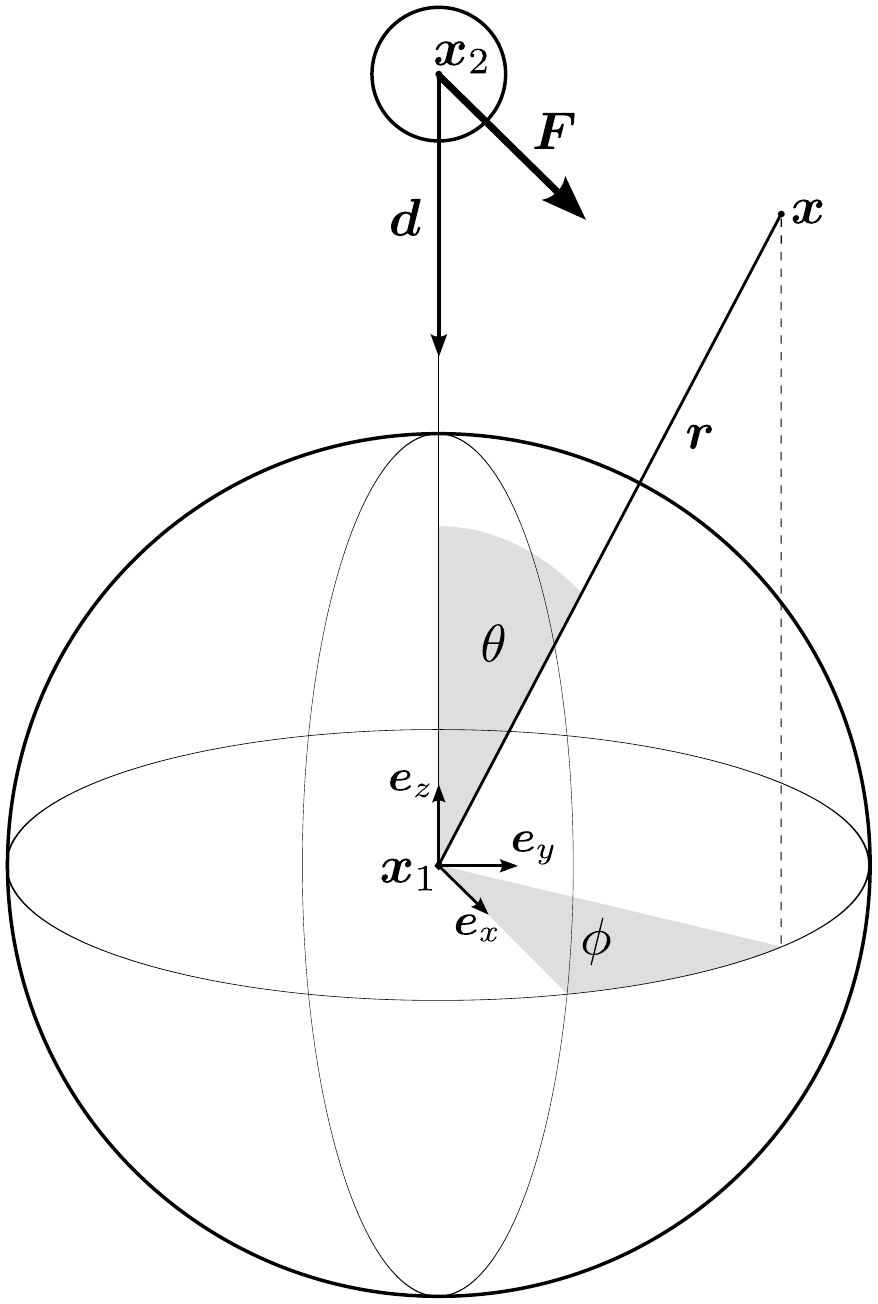}
\caption{The configuration of the system. A small solid particle of radius $b$ situated at $\xTwo = R \vect{e}_z$ nearby a large spherical capsule of undeformed radius $a$.
In an asymmetric situation, the force is directed perpendicularly to $\dd$ shown here along the $x$-direction.}
\label{illustration}
\end{center}
\end{figure}

\section{Singularity solution}\label{sec:singularitySolution}

We are interested in the flow field due to a point-force (Stokeslet) acting close to a large spherical capsule, for which we shall find a fully analytical solution. The Stokeslet is oriented perpendicularly to the line connecting its position and the center of the capsule. 
We introduce a spherical coordinate system, centered at the capsule position $\xOne$, with the point-force acting at  $\xTwo = R \vect{e}_z$.
The whole system is sketched in Fig.~\ref{illustration}.

Mathematically, the problem is reduced to solving the forced Stokes equations outside the capsule
\begin{align}
 \eta \Nabla^2 \vect{v} - \Nabla p + \vect{F} \delta (\X - \xTwo) &= 0 \, , \label{Stokes:Momentum} \\
 \Nabla \cdot \vect{v} &= 0 \, , \label{Stokes:Continuity}
\end{align} 
and homogeneous equations for the fluid inside
\begin{align}
 \eta \Nabla^2 \vect{v}^{(i)} - \Nabla p^{(i)}  &= 0 \, , \label{Stokes:Momentum_Inside} \\
 \Nabla \cdot \vect{v}^{(i)} &= 0 \, . \label{Stokes:Continuity_Inside}
\end{align}
Here $\vect{v}$ and $p$ denote the exterior velocity and pressure fields and the superscript $(i)$ stands for the corresponding interior fields.
For simplicity, we assume the fluid to have the same dynamic viscosity $\eta$ everywhere. The boundary conditions are imposed at $r=a$. 
We require the natural continuity of the fluid velocity field
 \begin{align}
  [v_\theta] &= 0 \, , \label{BC:v_phi} \\
  [v_\phi] &= 0 \, , \label{BC:v_the} \\
  [v_r] &= 0 \, , \label{BC:v_r} 
 \end{align}
and a fluid stress jump across the membrane imposed by its elastic properties, 
 \begin{align}
  [\sigma_{\theta r}] &= \Delta f_\theta^\mathrm{S} + \Delta f_\theta^\mathrm{B} \, , \label{BC:sigma_r_phi} \\
  [\sigma_{\phi r}] &= \Delta f_\phi^\mathrm{S} + \Delta f_\phi^\mathrm{B} \, , \label{BC:sigma_r_the} \\
  [\sigma_{rr}] &= \Delta f_r^\mathrm{S} + \Delta f_r^\mathrm{B} \, , \label{BC:sigma_r_r}
 \end{align}
where the notation $[w] := w(r = a^{+}) - w(r = a^{-})$ for the jump of a quantity $w$ across the membrane and the superscripts S and B denoting the shearing and bending related parts in the traction jump, respectively.
Throughout the remaining  of the paper, we scale all the lengths by the capsule radius $a$.
The corresponding quantities in physical units can be obtained by the transformation rules given in Appendix~B of part~I\cite{daddi17b}.
The components of the fluid stress tensor in spherical coordinates read \cite{kim13}
\begin{subequations}
 \begin{align}
   \sigma_{\theta r} &= \eta \left( v_{\theta,r} - \frac{v_\theta}{r} + \frac{v_{r,\theta}}{r} \right) \, , \label{sigma_r_phi} \\
   \sigma_{\phi r} &= \eta \left( \frac{v_{r,\phi}}{r \sin\theta} + v_{\phi,r} - \frac{v_\phi}{r} \right) \, , \label{sigma_r_theta} \\
   \sigma_{rr} &= -p + 2\eta v_{r,r} , \label{sigma_r_r}
 \end{align}
\end{subequations}
where the indices after commas indicate partial spatial derivatives, e.g. $v_{r,\phi} \equiv \partial{v_r}/\partial{\phi}$, etc. 

We model the elastic properties of the membrane by introducing its resistance towards shearing and bending. 
As derived in the Appendix, the linearized traction jumps due to shearing according to the Skalak model \cite{skalak73, Freund_2014}, characterized by a coefficient ${\lambda}$, in terms of the membrane deformation $\vect{u}$ read
\begin{subequations}
   \begin{align} \label{jump_u_The_Shear_main}
    \Delta f_\theta^\mathrm{S} &= 
     -\kS \bigg[ (2{\lambda}-1) u_{r,\theta} + {\lambda} u_{\theta,\theta\theta}    \\ \notag
            &+ {\lambda}u_{\theta,\theta} \cot\theta- u_\theta \left( {\lambda} \cot^2 \theta + {\lambda}-1 \right)  +\frac{u_{\theta,\phi\phi}}{2\sin^2\theta}  \\ \notag
            &  -\left({\lambda}+\frac{1}{2} \right) \frac{\cot\theta}{\sin\theta} \, u_{\phi,\phi} 
             +\left( {\lambda}-\frac{1}{2} \right) \frac{u_{\phi,\phi\theta}}{\sin\theta} \bigg]
            \, ,  \\ \label{jump_u_Phi_Shear_main}
   \Delta f_\phi^\mathrm{S} &= -\kS \bigg[
	    (2{\lambda}-1) \frac{u_{r,\phi}}{\sin\theta}+\left({\lambda}+\frac{1}{2} \right)\frac{\cot\theta}{\sin\theta} \, u_{\theta,\phi}  \\ \notag
 	    &+\left({\lambda}-\frac{1}{2} \right)\frac{u_{\theta,\phi\theta}}{\sin\theta} +\frac{1}{2} \left(1-\cot^2\theta \right) u_\phi   \\ \notag
	    &+ \frac{u_{\phi,\theta}}{2} \cot\theta +  \frac{u_{\phi,\theta\theta}}{2} + {\lambda} \, \frac{u_{\phi,\phi\phi}}{\sin^2\theta} \bigg] 
    \, ,  \\ \label{jump_u_R_Shear_main}
    \Delta f_r^\mathrm{S}  &= 
    \frac{2\kS}{3} (2{\lambda}-1) \left( 2u_r + u_{\theta,\theta} + u_{\theta} \cot \theta +\frac{u_{\phi,\phi}}{\sin\theta} \right) 
      \, ,  
   \end{align}
   \label{jump_Shear_main}
\end{subequations}
where ${\lambda} := C+1$ with $C$ being the ratio of the area expansion modulus $\kA$ and the shear modulus $\kS$ \cite{krueger11}.

The stress jump related to bending is derived from a linear isotropic model for the bending moments which is equivalent to the well-known Helfrich model \cite{helfrich73} for small deformations  \cite{guckenberger16, Guckenberger_preprint}.
The linearized traction due to bending reads (cf. Appendix)
\begin{subequations}
 \begin{align} \label{jump_u_The_Bending_main}
   \Delta f_\theta^\mathrm{B} &=  \kB \bigg[ \left(1-\cot^2\theta\right)u_{r,\theta} + u_{r,\theta\theta} \cot\theta  \\ \notag
   &+ u_{r,\theta\theta\theta}+(1+\cot^2\theta)\left( u_{r,\phi\phi\theta}-2 u_{r,\phi\phi}\cot\theta \right) \bigg]  \, ,  \\ \label{jump_u_Phi_Bending_main}
   \Delta f_\phi^\mathrm{B} &=  \kB  (1+\cot^2\theta) \big(  u_{r,\phi\theta} \cot\theta + 2u_{r,\phi} \\ \notag
   &+ u_{r,\phi\theta\theta} + (1+\cot^2\theta) u_{r,\phi\phi\phi}  \big) \sin\theta  \, ,  \\ \label{jump_u_R_Bending_main}
   \Delta f_r^\mathrm{B} &= \kB \bigg[ \left( 3\cot\theta+\cot^3\theta \right) u_{r,\theta} -  u_{r,\theta\theta}\cot^2\theta \\ \notag
   &+ 2 u_{r,\theta\theta\theta}\cot\theta + u_{r,\theta\theta\theta\theta}+  (1+\cot^2\theta) \big( 2u_{r,\phi\phi\theta\theta}  \\ \notag
  & - 2u_{r,\phi\phi\theta} \cot\theta + (1+\cot^2\theta) (4 u_{r,\phi\phi}+u_{r,\phi\phi\phi\phi}) \big) 
  \bigg]  \, ,
 \end{align}
 \label{jump_Bending_main}
\end{subequations}
where $\vect{u}(\theta, \phi)$ is the membrane displacement field.
These expressions reduce to the axisymmetric case of part I by setting $u_\phi=0$ and dropping all $\phi$-derivatives.
The displacement is related to the fluid velocity at $r=1$ via the no-slip condition, 
\begin{equation}
\left. \vect{v} \right|_{r = 1} = \td{\vect{u}}{t} \, , \notag
\end{equation}
which in the Fourier space takes the form 
\begin{equation}
\left. \vect{v} \right|_{r = 1} = i\omega \, \vect{u} \, . \label{no-slip-equation}
\end{equation}

Our approach is inspired by the work of Fuentes \textit{et al.} \cite{fuentes89}, who computed  the solution of the Stokes equation nearby a viscous drop for the asymmetric force case. We write the exterior fluid velocity outside the capsule as
\begin{equation}
 \vect{v} = \vStok + \vect{v}^{*} \, , \notag
\end{equation}
where $\vStokcom_i := \G_{ij} (\vect{x} - \xTwo) {F}_j $ is the velocity field induced by a point-force acting at $\xTwo$ in an infinite fluid, and $\vect{v}^{*}$ is the flow due to an image system required to satisfy the boundary conditions at the capsule membrane, also called the reflected flow.

Now we sketch briefly the main steps of our solution methodology.
Firstly, we express the Stokeslet velocity $\vStok$  at $\xTwo$ in terms of spherical harmonics, which are then transformed into harmonics centered at $\xOne$ via the Legendre expansion. 
Secondly, we write a multipole expansion for the image system $\vect{v}^{*}$ at $\xOne$, and afterward we rewrite it in terms of spherical harmonics based at~$\xOne$.
Thirdly, the solution inside the capsule $\vect{v}^{(i)}$ is written using Lamb's general solution \cite{lamb32}, also expressed in terms of spherical harmonics at $\xOne$.
The last step consists of determining the unknown series expansion coefficients by satisfying the boundary conditions at the membrane stated by Eqs.~\eqref{BC:v_phi} through~\eqref{BC:sigma_r_r}.

In conjunction with the results of part~I on the axisymmetric motion, the general solution of the Stokes equations for an arbitrary force direction is thus obtained.


\subsection{Stokeslet representation}

We begin with writing the Stokeslet positioned at $\xTwo$,
\begin{equation}
 \vStok = \Gmatr \cdot\F = \frac{1}{8\pi\eta} \left(\frac{\boldsymbol{1}}{s} + \vect{s} {\Nabla_2} \frac{1}{s} \right)\cdot\F , \label{stokeslet_at_X2}
\end{equation}
where $\vect{s} := \vect{x}-\xTwo$ and $s:= |\vect{s}|$.
Here ${\nabla_2}_j := {\partial }/{\partial  {x_2}_j}$ denotes the gradient operator taken with respect to $\xTwo$.
Using the Legendre expansion, the harmonics based at $\X_2$ can be expressed in terms of those centered at $\X_1$ as
\begin{equation}
 \frac{1}{s} = \infSum \frac{r^{2n+1}}{R^{n+1}} \frac{(\vecD \cdot \Nabla)^n}{n!} \frac{1}{r} \, , \notag
\end{equation}
with the unit vector $\vecD := ( \xOne - \xTwo)/R = -\vect{e}_z$, $\R= \X-\X_1$, and $r := |\R|$. The derivatives with respect to $\X_2$ are taken care of by noting that
\begin{equation}
 \Nabla_2 \frac{1}{R^{n+1}} = \frac{n+1}{R^{n+2}}\,  \vecD \, , \qquad  (\vecD \cdot \Nabla_2)\, \vecD = 0 \, . \notag
\end{equation}
Moreover, we denote by $\vp_n$ the harmonic of degree $n$, related to the Legendre polynomials of degree $n$, $P_n$ by \cite{abramowitz72}
\begin{equation}
 \vp_n (r, \theta) := \frac{(\dd \cdot \Nabla)^n}{n!} \frac{1}{r} = \frac{1}{r^{n+1}} P_n (\cos \theta) \, . \notag
\end{equation}
In this work, we focus our attention on the asymmetric case when the force is purely tangential and therefore $\vect{F}\cdot\vect{d}=0$. Taking this into account, the Stokeslet in Eq.~\eqref{stokeslet_at_X2} can be written as 
\begin{equation}
 \begin{split}
  8\pi\eta \vStok &= \F \infSum \frac{r^{2n+1}}{R^{n+1}} \, \vp_n 
  -\R \infSumOne \frac{r^{2n+1}}{R^{n+2}} \left(\F\cdot\Nabla\right)\vp_{n-1} \\
  &-\dd \infSumOne \frac{r^{2n+1}}{R^{n+1}} \left(\F\cdot\Nabla\right)\vp_{n-1} \, .
 \end{split} \notag
\end{equation}
Thus we have expressed the Stokeslet solution in terms of spherical harmonics centered at $\X_1$.
By defining $\vecT = \F \times \vecD$, we have the identity
\begin{equation}
 \vecD (\F\cdot\bNabla) \varphi_n = (\vecT \times \bNabla) \varphi_n + (n+1) \F  \varphi_{n+1} \, . \label{identityRotlet}
\end{equation}
Moreover, for $\F\cdot\dd=0$, we can write
\begin{align} 
\label{eliminateFPhi}
(2n+3)\R\psi_n &= - {r^2} \Nabla\psi_n+\Nabla\psi_{n-2} \\ \nonumber 
&-(2n+1)(n+1)\F\vp_{n} - (2n+1)\bgamma_{n-1} \, ,  
\end{align}
where we have defined
\begin{equation}
\psi_n = (\F\cdot\Nabla)\vp_{n} \, , \qquad \bgamma_n = (\T\times\Nabla)\vp_{n} \, . \notag
\end{equation}
Note that the harmonics $\psi_n$ are defined differently than in part~I and that the additional set $\bgamma_n$ is not required for the simpler axisymmetric case of part~I.
Finally, the Stokeslet can be written as
\begin{equation}
 \begin{split}
  8\pi\eta\vStok &= \infSumOne \left[\frac{n-2}{(2n-1){n}}\frac{{r^{2n+1}}}{R^n} \right. \\
                 &-\left. \frac{n}{(n+2)(2n+3)}\frac{{r^{2n+3}}}{R^{n+2}}\right]\Nabla\psi_{n-1} \\ 
                 &-\frac{2}{n+1} \frac{{r^{2n+1}}}{R^{n+1}} \, \bgamma_{n-1} \\
                 &+ \left[\frac{(n-2)(2n+1)}{n(2n-1)}\frac{{r^{2n-1}}}{R^n} - \frac{{r^{2n+1}}}{R^{n+2}}\right]\R\psi_{n-1}. \label{Stokeslet_finalize}
 \end{split}
\end{equation}
We have chosen the vector basis functions here to be $\Nabla\psi_n$, $\vect{r}\psi_n$, and $\bgamma_n$. We now proceed to deriving analogous expansions for the reflected flow and the velocity inside the capsule, in order to finally match them using the boundary conditions given above.

\subsection{Image system representation}

The corresponding image system representation can be written as a multipole expansion, which involves the derivatives of the free-space Green's function $\Gmatr(\vect{r})$, as\cite{kim13}
\begin{align} 
\label{imageSystem_init}
8\pi\eta \bv^* &= \infSum \left[ A_n\frac{(\dd\cdot\Nabla)^n}{n!} \, \Gmatr(\R)\right.  \nonumber  \\ 
&+\left.B_n\frac{(\dd\cdot\Nabla)^n}{n!}  \nabla^2\Gmatr(\R) \right] \cdot \F \\ \nonumber
&+ \infSum \left[ C_n\frac{(\dd\cdot\Nabla)^n}{n!}  (\T\times\Nabla)\frac{1}{r}\right] \, . 
\end{align}

We convert these expressions into harmonics $\varphi_n$ using the identity
\begin{equation}
\frac{(\dd\cdot\Nabla)^n}{n!} \,\G_{ij}(\R) = \delta_{ij}\vp_n - r_i \nabla_j \vp_{n} - d_i\nabla_j \vp_{n-1}, \notag
\end{equation}
and the fact that the Laplacian of the Oseen tensor is written conveniently as
\begin{equation}
\nabla^2 \Gmatr (\R) = -2 \Nabla\Nabla \frac{1}{r} \, . \notag
\end{equation}
Making use of Eq.~\eqref{identityRotlet}, the image system solution can finally be written as
 \begin{align}
  8\pi\eta\bv^* &= \infSum \bigg[A_n\Bigl((1-n)\F\vp_{n} - \R\psi_{n}\Bigr)\nonumber \\ 
&-  2B_n \Nabla\psi_{n} \bigg] + \infSumOne \left[C_n - A_{n+1} \right] \bgamma_{n} \, .
  \label{imageSystem_finalized}
 \end{align}

\subsection{The interior solution}
 The interior solution has a generic form derived first by Lamb\cite{lamb32,kim13}. 
 It involves three families of unknown coefficients and can be written in the asymmetric situation as
  \begin{align}
   \label{insideSolution_finalized}
  8\pi\eta\bv^{(i)} &=  \infSumOne c_n \bigg[ r^{2n-1} \bgamma_{n-1} 
		    + (2n-1) r^{2n-3} (\T\times\R) \vp_{n-1}\bigg] \notag \\ 
		    &+ b_n \left[ \frac{r^{2n+1}}{n}\Nabla\psi_{n-1} + \frac{2n+1}{n} r^{2n-1}\R \psi_{n-1} \right] \notag \\ 
		    &+a_n \bigg[\frac{n+3}{2n} r^{2n+3} \Nabla\psi_{n-1}  \\ 
		    &+ \frac{(n+1)(2n+3)}{2n} r^{2n+1} \R \psi_{n-1}\bigg] \, . \notag 
  \end{align}
 We note that the interior solution here has three unknown coefficients while the axisymmetric motion in part~I involves only two.
 
 \subsection{The full flow field}
 
 The velocity fields $\vStok$, $\vect{v}^*$, and $\bv^{(i)}$ thus suffice to describe the flow in the whole space. 
{ The matching conditions at the surface of the capsule are determined by the known stress jump due to the membrane elasticity and continuity of the velocity field, as expressed by Eqs.~\eqref{BC:v_phi} through \eqref{BC:sigma_r_r}.
 These allow computation of the free constants ($A_n$, $B_n$, $C_n$ for the reflected flow, and $a_n$, $b_n$, $c_n$ for the inner flow) as detailed in  Appendix~\ref{appendix:determination}.}


\section{Particle self-mobility}\label{sec:particleSelfMobility}

In the preceding section, we have computed the Green's function for the problem of a point-force acting in the direction tangential to the surface of an elastic spherical capsule. 
The exterior velocity field due to a Stokeslet is then given by $\vStok +\vect{v}^*$. 
In this section, we discuss the consequences of the presence of the membrane for the motion of the nearby particle. In order to assess the effects of the presence of the capsule, we now compute the leading-order correction term to the particle self-mobility.
We assume an external force $\vect{F}_2$ to be acting on the solid particle and no force or torque to be exerted on the capsule.

The zeroth-order solution for the particle velocity is $\vect{V}_2^{(0)} = \mu_0 \vect{F}_2$ as given by the Stokes law with $\mu_0 := 1/(6\pi\eta b)$ being the usual bulk mobility.
The leading-order correction to the particle self-mobility is computed from the image solution as
\begin{equation}
 \vect{v}^* \big|_{\X = \X_2} = \Delta \mu \vect{F}_2 \, .
\end{equation}
Making use of the following relations
 \begin{align}
  \frac{(\vecD\cdot\bNabla)^n}{n!} \Gmatr (\vect{x}-\xOne) \bigg|_{\vect{x}=\xTwo}\cdot  \Fext &= \frac{1}{R^{n+1}} \Fext \, , \notag \\
  \frac{(\vecD\cdot\bNabla)^n}{n!} \bNabla^2 \Gmatr (\vect{x}-\xOne) \bigg|_{\vect{x}=\xTwo} \cdot\Fext &= \frac{(n+1)(n+2)}{R^{n+3}} \Fext \, , \notag \\ \notag
  \frac{(\vecD\cdot\bNabla)^n}{n!}  \left( \T \times \bNabla \right) \frac{1}{r} \bigg|_{\vect{x}=\xTwo} &= -\frac{n+1}{R^{n+2}} \Fext \, ,
 \end{align}
together with Eq.~\eqref{imageSystem_init}, the scaled particle self-mobility function reads
\begin{equation}
 \frac{\Delta \mu}{\mu_0} = \frac{3b}{4} \infSum \big[ A_n + (n+1)(n+2) \xi^2 B_n  - (n+1) \xi C_n  \big] \xi^{n+1} \, ,  \label{mobilityCorrection}
\end{equation}
where $\xi := 1/R \in [0,1)$.
We denote by $f_n (\xi)$ the general term of the function series giving the particle scaled mobility correction stated above.
For large $n$, we obtain the leading order asymptotic behavior
\begin{equation}
 f_n(\xi) = \frac{3b}{16} \left( 1-\xi^2 \right)^2 n^2 \xi^{2n+4} + \bigO \left( n \xi^{2n} \right) \, , \label{f_n} 
\end{equation}
which is independent of shearing and bending properties.
The number of terms required for convergence can thus be estimated for a given precision as in Appendix C of part~I\cite{daddi17b}.

\begin{figure}
 \includegraphics[scale=1]{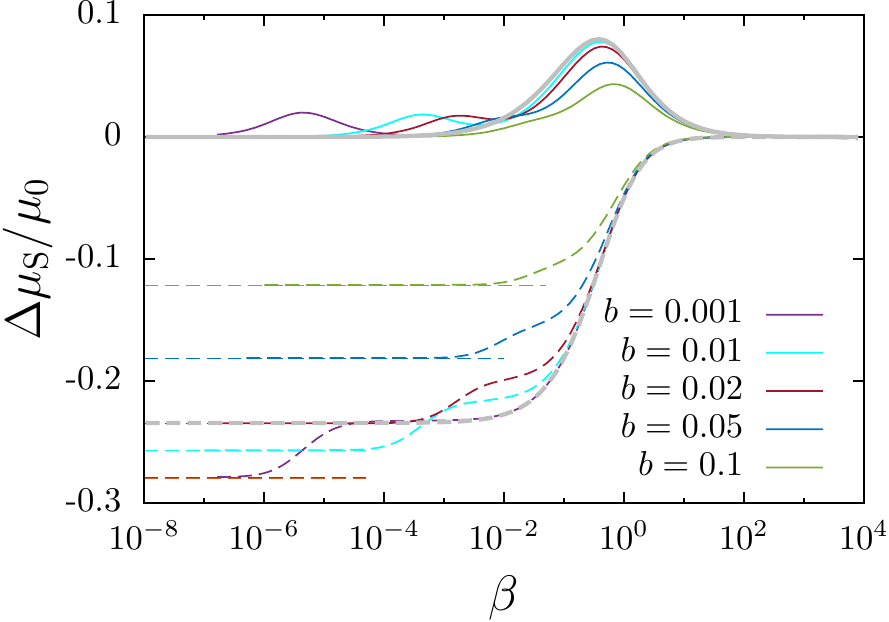}
 \caption{(Color online) Scaled particle self-mobility correction versus scaled frequency $\beta$ for various values of the small particle radius $b$ for a membrane with pure shearing. 
 The real and imaginary  parts are shown as dashed and solid lines, respectively.
 Dashed lines on the vertical axis at small $\beta$ represent the hard-sphere limit given by Eq.~\eqref{hardSphere}.
 The curve in gray corresponds to the self-mobility correction for a planar membrane given by Eq.~\eqref{mobilityCorrection_Planar_Shearing}.
 Here the solid particle is set at $h=2b$. }
 \label{sphericalParaShearing_A_Effect_Planar}
\end{figure}

\begin{figure}
 \includegraphics[scale=1]{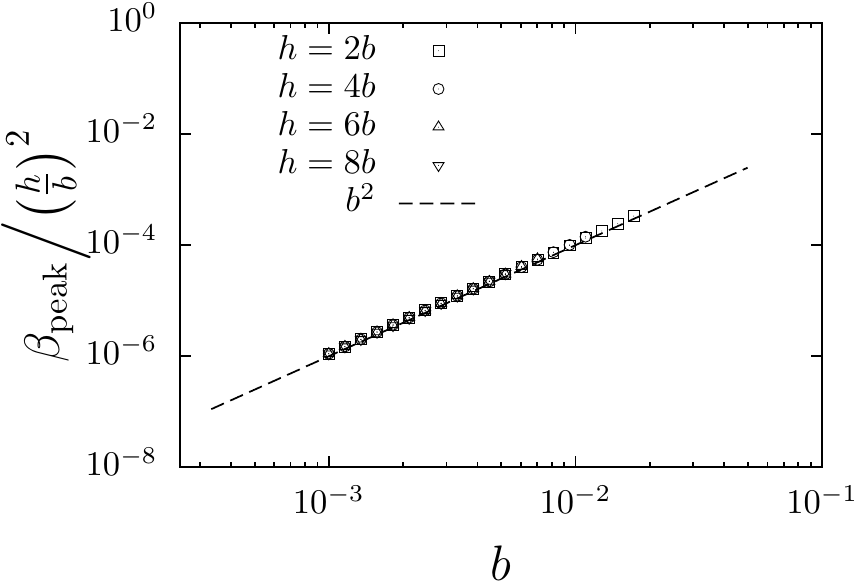}
 \caption{ Log-log plot of the rescaled peak-frequency ${\beta}_\mathrm{peak}$ versus $b$ for different particle-to-membrane distance $h$.
 }
 \label{betaPeak}
\end{figure}

It is worth to mention here that for finite membrane shearing modulus (i.e. for a non hard-sphere), no net force is exerted on the spherical capsule, since $A_0 = 0$. In this case, the capsule is also torque-free, since $C_0 - A_1 = 0$.
For a hard-sphere, however, additional singularities should be involved in the computation of particle mobility to ensure the force- and torque-free assumptions (see Fuentes \textit{et al.} \cite{fuentes89} for further details.)

\subsection{Shearing contribution}

For a membrane exhibiting a shearing-only resistance, the self-mobility correction can be computed by plugging the expressions of $A_n$, $B_n$ and $C_n$ as stated respectively by Eqs.~\eqref{A_n_Shearing}, \eqref{B_n_Shearing} and \eqref{C_n_Shearing} into Eq.~\eqref{mobilityCorrection}.
By taking the limit $\alpha\to\infty$ we recover the rigid sphere limit, 
\begin{equation}
 \frac{\Delta \mu_{\mathrm{S}, \infty}}{\mu_0} := \lim_{\alpha\to\infty}  \frac{\Delta \mu}{\mu_0} = - \frac{\xi^5 (17 + \xi^2)}{16(1-\xi^2)} \frac{b}{R} \, ,
 \label{hardSphere}
\end{equation}
in agreement with the result by Ekiel-Je{\.z}ewska and Felderhof \cite[Eq. (2.26)]{ekiel15}.
Taking in addition an infinite membrane radius, we obtain 
\begin{equation}
 \frac{\Delta \mu_{\mathrm{S}, \infty}}{\mu_0} = -\frac{9}{16} \frac{b}{h} \, , \label{hardSpherePlanar}
\end{equation}
where $h := R- 1 $ is the distance separating the center of the solid particle to the closest point on the capsule surface.
We therefore recover the leading-order mobility correction for the motion parallel to a planar hard-wall as computed by Lorentz \cite{lorentz07}.

To characterize the dynamic effects at different forcing frequencies, we define the dimensionless shearing frequency as $\beta:=6B \eta \omega h/\kS$ where $B:=2/{\lambda}$. In Fig.~\ref{sphericalParaShearing_A_Effect_Planar} we show the scaled self-mobility correction for a membrane with pure shearing with $C=1$ ($\lambda=2$) versus the scaled frequency $\beta$ upon variation of the particle radius $b$ while keeping $h=2b$. 
We remark that the real part of the mobility correction (shown as dashed lines) is an increasing function of frequency while the imaginary part (shown as solid lines) has the typical peak structure attributed to the memory effect induced by membrane elasticity.
In the vanishing frequency limit, the mobility correction near a hard-sphere given by Eq.~\eqref{hardSphere} is recovered. 

As the particle radius decreases, we observe that in the high frequency regime both the real and imaginary parts follow faithfully the evolution predicted for a planar membrane which reads \cite{daddi16}
\begin{align} \label{mobilityCorrection_Planar_Shearing}
 \frac{\Delta \mu_\mathrm{S}}{\mu_0} &= \frac{3}{8} \frac{b}{h} \bigg[ -\frac{5}{4}+\frac{\beta^2}{8} + {i\lambda\beta } e^{i\lambda \beta} \E_1 \left( {i\lambda\beta} \right) \\ \nonumber 
 & -\frac{3i\beta}{8}+ \left( -\frac{\beta^2}{2}+\frac{i\beta}{2} \left( 1-\frac{\beta^2}{4} \right) \right) e^{i\beta} \E_1(i\beta) \bigg] \, . 
\end{align}
The peak occurring at $\beta\sim 1$ can be estimated by a balance between membrane elasticity and fluid viscosity as $\omega\sim \kS/(\eta h)$.
This peak is attributed to membrane in-plane displacements $u_\theta$ and $u_\phi$ which are observed in the same way for planar membranes.
The second peak of small amplitude occurring in the low frequency regime is attributed to membrane normal displacement along $u_r$ which is not involved in the traction jumps for planar membranes.
In fact, for the axisymmetric motion examined in part~I \cite[see][Fig.~2]{daddi17b}, we observe that the low-frequency peak has a remarkably higher amplitude since the membrane displacement $u_r$ manifests itself in a more pronounced way for the motion perpendicular than for the motion parallel to the membrane.

Interestingly, the frequency corresponding to the left peak of the imaginary part of the mobility correction is found to be proportional to $b^2$, as plotted in Fig. \ref{betaPeak}. 
For different radii and separations, the same master curve is recovered and the second peak frequency position can conveniently be estimated from the relation $\beta_\mathrm{peak} = h^2$.
It is worth noting that a scaling relation $\beta_\mathrm{peak} = 2h^2$ has been obtained for the axisymmetric motion considered in part~I.

\begin{figure}
 \includegraphics[scale=1]{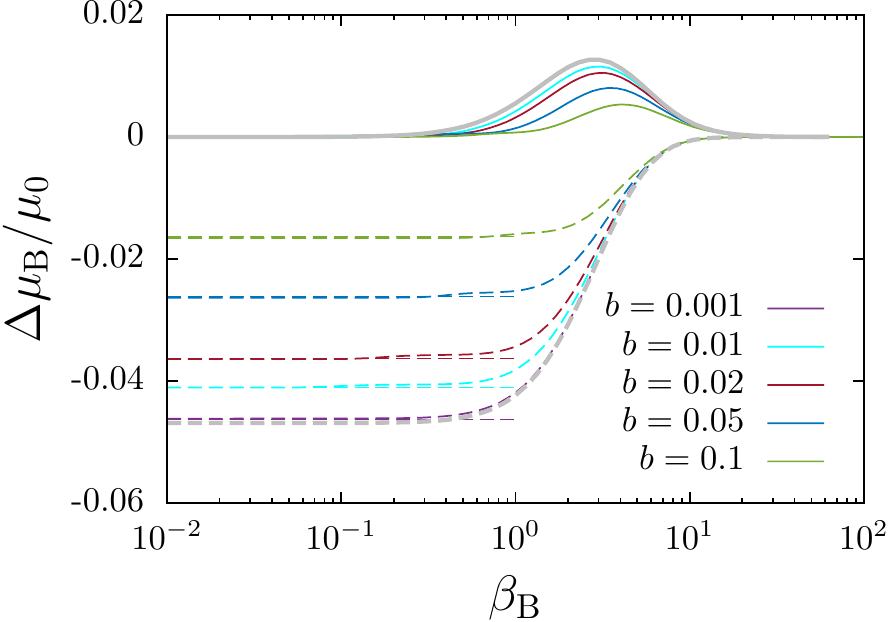}
 \caption{(Color online) Scaled self-mobility correction versus $\betaB$ for various values of the capsule radius, for a membrane with pure bending.
 The dashed and continuous lines represent the real and imaginary parts, respectively.
 The horizontal dashed lines are the vanishing frequency limits approximated by Eq.~\eqref{bendingLimit_Approximation}.
 The curve shown in gray is the solution given by Eq.~\eqref{mobilityCorrection_Planar_Bending} for a planar membrane.
 Here we take $h=2b$.
 }
 \label{sphericalParaBending_A_Effect_Planar}
\end{figure}

\subsection{Bending contribution}

For a membrane possessing only bending rigidity, the self-mobility correction is determined by plugging the expressions of  $A_n$, $B_n$ and $C_n$ as stated respectively by Eqs.~\eqref{A_n_Bending}, \eqref{B_n_Bending} and \eqref{C_n_Bending} into Eq.~\eqref{mobilityCorrection}.
By taking the limit $\alphaB\to\infty$, the leading-order self-mobility can be approximated by
\begin{align} \label{bendingLimit_Approximation}
  \frac{\Delta \mu_{\mathrm{B}, \infty}}{\mu_0} :=& \lim_{\alphaB\to\infty}  \frac{\Delta \mu}{\mu_0} \simeq -\frac{\xi^5}{70(1-\xi^2)} \times \\ \nonumber
  &\left[ -9+71 \xi^2 -\frac{183}{2} \xi^4 + \frac{341}{8} \xi^6 \right] \frac{b}{R} \, , 
\end{align}
which for an infinite radius reads
\begin{equation}
  \frac{\Delta \mu_{\mathrm{B}, \infty}}{\mu_0} = -\frac{3}{32} \frac{b}{h} \, , \notag
\end{equation}
corresponding to the vanishing frequency limit for an idealized membrane with pure bending.
Note that this limit is different from the hard-sphere limit but identical to that of a planar interface separating two immiscible fluids having the same viscosity \cite{lee79, happel12}.
A similar behavior has been observed for planar membranes with pure bending resistance.
This can be justified by the fact that the bending traction jump stated by Eq.~\eqref{jump_Bending_main} is determined only by the radial displacement $u_r$ and does not involve the tangential displacements $u_\theta$ and $u_\phi$.
As a result, even for an infinite bending modulus, the membrane tangential displacements remain completely free.
This behavior is in contrast to the hard-sphere where all the displacement field components are restricted.

We define the characteristic frequency for bending as $\betaB := 2h(4\eta\omega/\kB)^{1/3}$. 
In Fig.~\ref{sphericalParaBending_A_Effect_Planar}, we show the scaled self-mobility correction versus $\betaB$ of a particle moving nearby a membrane with bending-only resistance.
Unlike a membrane with pure shearing resistance, the second low frequency peak is not observed nearby a membrane with pure bending resistance.
This can be understood since the traction jumps due to bending involve only the radial displacement $u_r$.
The peak observed at $\betaB^3 \sim 1$ is the characteristic peak for bending which can be estimated by a simple balance between membrane bending and fluid viscosity as $\omega \sim \kB/(h^3 \eta)$.
This trend is in contrast to what has been observed for membrane with pure shearing resistance where the traction jumps involve both the radial and tangential displacements.

For sufficiently small values of $b$, we observe that both the real and imaginary parts of the mobility correction are in good agreement with the planar membrane solution \cite{daddi16}
\begin{equation}
 \frac{\Delta \mu_\mathrm{B} (\betaB)}{\mu_0} = \frac{3}{64} \frac{b}{h} \left[ -2 + \frac{i\betaB^3}{3} \left( \phi_{+} + e^{-i\betaB} \E_1 (-i\betaB) \right) \right] \, , \label{mobilityCorrection_Planar_Bending}
\end{equation}
wherein
\begin{equation}
 \phi_+ = e^{-i \overline{\zB}} \E_1(-i \overline{\zB}) + e^{-i\zB} \E_1(-i\zB) \, ,  \notag
\end{equation}
and $\zB = \betaB e^{2i\pi/3}$. 
As a result, a very good estimate of particle mobility can be made for large capsules with bending-only resistance from the planar membrane limit.
For moderate and small capsule radii however, the planar membrane solution leads to a reasonable agreement only in the high frequency regime for which $\betaB > 1$, in the same way as observed in part~I for the motion perpendicular to the membrane \cite{daddi17b}.

\subsection{Shearing-bending coupling}

\begin{figure}
 \includegraphics[scale=1]{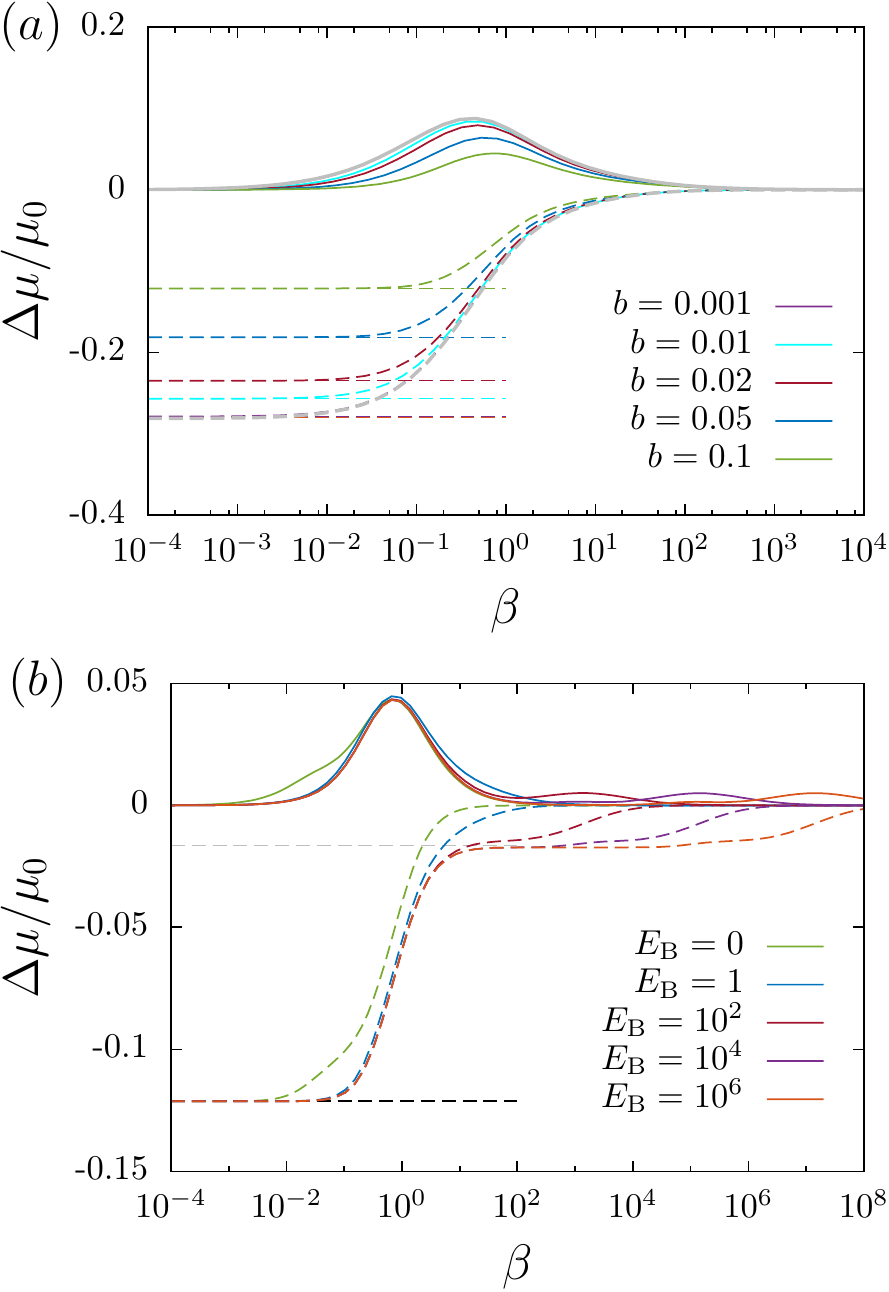}
 \caption{(Color online) $a)$ Scaled particle mobility correction versus $\beta$ for various values $b$ for a membrane possessing both shearing and bending rigidities. 
 The real and imaginary parts are shown as dashed and solid lines, respectively.
 Horizontal dashed lines are the hard-sphere limit given by Eq.~\eqref{hardSphere}.
 The curve shown in gray corresponds to the mobility correction for a planar elastic membrane \cite{daddi16} as obtained by linear superposition of Eqs.~\eqref{mobilityCorrection_Planar_Shearing} and \eqref{mobilityCorrection_Planar_Bending}.
 Here the solid particle is set at $h=2b$ and the membrane has a reduced bending modulus $\EB=1$.
 $b)$ Scaled mobility correction versus $\beta$ for various values of $\EB$.
 The horizontal dashed line in black is the hard-sphere limit given by Eq.~\eqref{hardSphere}, whereas the gray dashed line corresponds to the infinite bending rigidity limit predicted for a bending-only membrane as given by Eq.~\eqref{bendingLimit_Approximation}.
 Here we take  $b=1/10$ and  $h=2b$.
 }
 \label{Fig_EB}
\end{figure}

Unlike for a planar membrane, shearing and bending are intrinsically coupled for a spherical membrane. 
As a result, the mobility correction is not a linear superposition of independent contributions from shearing and bending.
A similar coupling behavior is observed between two planar elastic interfaces \cite{daddi16b} or thermally fluctuating membranes \cite{auth07, kovsmrlj14}.
In order to investigate this coupling, we define the reduced bending modulus as $\EB:=\kB/(\kS h^2)$, a parameter that quantifies the relative contributions of shearing and bending.

In Fig.~\ref{Fig_EB}~$a)$ we show the scaled self-mobility correction versus $\beta$ nearby a membrane with both shearing and bending resistances upon varying $b$.
We observe that in the high frequency regime, i.e. for $\beta > 1$, the mobility correction follows the evolution predicted for a planar elastic membrane.
For lower values of $b$, the planar membrane prediction leads to a very good estimation even deeper into the low-frequency regime.
In the following, we take $h=2b$ and a membrane reduced bending modulus $E_\mathrm{B}=1$, for which shearing and bending effects manifest themselves equally.

In Fig.~\ref{Fig_EB}~$b)$, we show the particle self-mobility correction versus the scaled frequency $\beta$ for a membrane with both shearing and bending rigidities upon varying the reduced bending modulus $E_\mathrm{B}$ while keeping $b=1/10$ and $h=2b$.
For $E_\mathrm{B} = 0$ (shearing-only membrane), a low frequency peak as in Fig.~\ref{sphericalParaShearing_A_Effect_Planar} emerges.
For higher values of $\EB$ this peak completely disappears confirming our hypothesis that it is due to radial deformations which are suppressed by bending resistance.

The imaginary part exhibits a high frequency peak of typically constant height for increasing $\EB$.
Since $\beta$ and $\beta_\mathrm{B}$ are related by 
\begin{equation}
 \betaB^3 = \frac{16}{3B \EB} \, \beta \, , \notag
\end{equation}
the peak observed at $\beta \sim 1$ is attributed to shearing, whereas the peak occurring in the high frequency regime is attributed to bending, since $\beta \sim \EB$ for $\betaB^3 \sim 1$.
Particularly, for $E_\mathrm{B} = 1$, the  position of the two peaks coincides as $\beta \sim \beta_\mathrm{B}^3$ for which shearing and bending effects have equal contribution.

\section{Diffusion nearby cell membranes}\label{sec:diffusion}

{
The analytical predictions of the particle self-mobility presented in the previous section serve as a basis for the study of particle diffusional motion nearby spherical cell membranes.
The determination of the mean-square displacement (MSD) requires as an intermediate step the computation of the velocity autocorrelation function which is derived from the fluctuation-dissipation theorem as \cite{kubo66, kubo85}
\begin{equation}
 \phi_{v} (t) = \frac{\kBolt T}{2\pi} \int_{-\infty}^\infty \left( \mu(\omega) + \overline{\mu(\omega)} \right) e^{i\omega t} \, \Intd \omega \, , 
\end{equation}
wherein $\kBolt$ is the Boltzmann constant and $T$ is the absolute temperature of the system. 
In this way, the particle MSD is computed as
\begin{equation}
 \langle \Delta r (t)^2 \rangle = 2 \int_0^t (t-s) \phi_v (s) \, \Intd s \, .
\end{equation}
Further, for the sake of convenience, we define the excess MSD as the membrane induced \emph{scaled} correction to the full MSD as   \cite{daddi16}
\begin{equation}
 \Delta (t) := 1 - \frac{\langle \Delta r (t)^2 \rangle}{2D_0 t} \, , \label{excessMSD}
\end{equation}
wherein $D_0 = \mu_0 \kBolt T$ is the usual bulk diffusion coefficient predicted from Einstein theory \cite{einstein05, einstein06}.
}

{
In typical physiological situations, red blood cell membranes have a shear modulus $\kS=5\times 10^{-6}$N/m, a bending modulus $\kB=2\times 10^{-19}$Nm \cite{Freund_2014} and a discocyte shape of local radius $a=10\mu$m.
We then consider a solid particle of radius $b=a/10$ located a distance $h=2b$ for which the reduced bending modulus $\EB = 1$.
We scale the time by the characteristic time scale for shearing $\TS = 6h\eta / \kS$ which is of about 0.3~$\mu$s considering a typical dynamic viscosity of blood plasma $\eta = 1.2~$mPas.}

\begin{figure}[t]
 \includegraphics[scale=1]{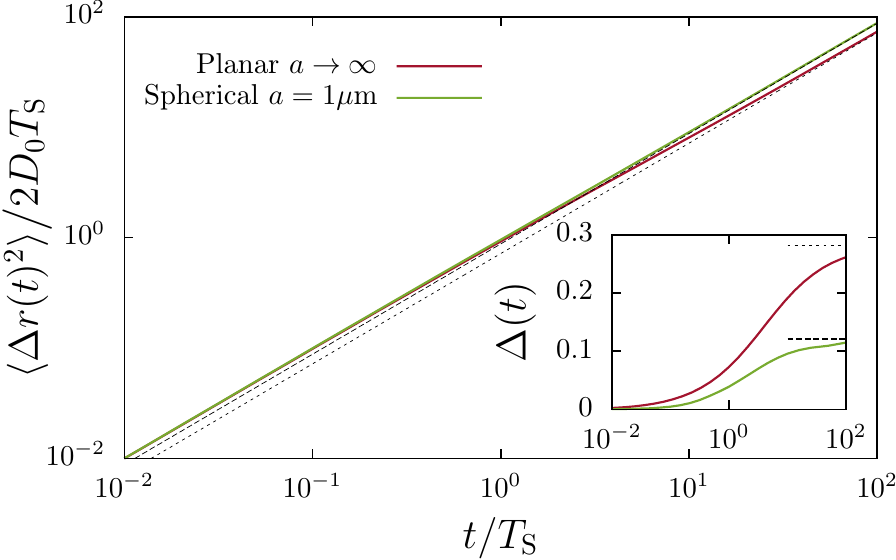}
 \caption{{ Mean-square displacement versus time for Brownian motion of a 100nm particle parallel to a planar and a spherical red blood cell membrane with curvature radius $a=1\mu$m. 
 Dotted and dashed lines represent the corresponding MSDs near a hard wall or sphere, respectively.
 The inset shows the variation of the excess MSD as defined by Eq.~\eqref{excessMSD}.
}}
 \label{diffusionSphMemParallel}
\end{figure}

{
In Fig.~\ref{diffusionSphMemParallel}, we show the scaled MSD versus the scaled time for a particle diffusing nearby a planar or a spherical membrane using the above mentioned parameters.
We observe that at short time scales of motion, the MSD follows a linear bulk behavior and the corresponding excess MSD amounts to very small values. 
This behavior is justified by the fact that the particle does not yet perceive the presence of the membrane at these very short time scales.
As the time increases, the effect of the membrane becomes noticeable and the particle experiences at intermediate time scales a long-lasting subdiffusive regime that can extend up to $10^2~\TS$ nearby a spherical membrane and even further for a planar membrane.
In the steady limit in which $t \gg \TS$, the MSD progressively approaches the value predicted nearby a hard boundary.
For the current set of physically realistic parameters, the steady excess MSD is found to be about twice larger for a planar membrane than that for a spherical membrane.
Therefore, accounting for membrane curvature becomes crucial for an accurate and precise computation of the particle diffusional motion.
}

\section{Capsule motion and deformation}\label{sec:capsuleMotionAndDeformation}

Having analyzed the capsule-induced correction to the self-mobility, we now focus on the motion of the capsule itself. This is characterized by the pair-mobility $\mu^{12}$, defined as the ratio between the velocity of the capsule centroid $V_1$ and the force $F_2$ applied on the nearby solid particle such that $V_1 = \mu^{12} F_2$. Without loss of generality, we assume that the force is directed along the $x$~direction. 
The capsule translational velocity can be computed by volume integration of the fluid velocity inside the capsule \cite{felderhof14},
\begin{equation}
 V_1 (\omega) = \frac{1}{\Omega} \int_0^1 \avg{v_x^{(i)} (r,\phi,\theta,\omega)} \, r^2  \, \Intd r  , \notag
\end{equation}
wherein $\Omega :=4\pi/3$ is the volume of the undeformed capsule, $\avg{\cdot}$ denotes angular averaging defined by Eq.~\eqref{averagingDefinition}, and $v_x^{(i)} = v_r^{(i)} \sin \theta \cos \phi + v_\theta^{(i)} \cos \theta \cos \phi - v_\phi^{(i)} \sin \phi$.
After integration, only the terms with $n=1$ of the series remain, leading to the frequency-dependent pair-mobility
\begin{equation}
 \mu^{12} = -\frac{1}{8\pi\eta} (a_1+b_1-c_2) \, ,  \notag
\end{equation}
which after computation simplifies to
\begin{equation}
 6\pi\eta \mu^{12} = \frac{3}{4} \, \xi + \frac{\xi^3}{4} \frac{3 + (2{\lambda}-1) \alpha}{5 + (2{\lambda}-1) \alpha} \, . \label{mobilityCorrection_Pair}
\end{equation}

The correction to the pair-mobility can therefore be expressed as a Debye-type model with a relaxation time given by
\begin{equation}
 \tau = \frac{15}{2(2{\lambda}-1)} \frac{\eta}{\kS} \, , \notag
\end{equation}
which is identical to that obtained for the axisymmetric motion \cite{daddi17b}.

In the limiting cases, two known results are recovered.
Firstly, for $\alpha \to \infty$, we obtain the leading-order pair-mobility between two unequal hard-spheres 
\begin{equation}
 \lim_{\alpha \to \infty} 6\pi\eta \mu^{12} = \frac{3}{4} \, \xi + \frac{\xi^3}{4} \, . \label{pairMobi_hardSphere}
\end{equation}
Secondly, for $\alpha \to 0$, we get the leading-order pair-mobility between a solid particle and large spherical viscous drop
\begin{equation}
 \lim_{\alpha \to 0} 6\pi\eta \mu^{12} = \frac{3}{4} \, \xi + \frac{3}{20} \, \xi^3 \, , \label{pairMobi_viscousDrop}
\end{equation}
both of which are in agreement with the results by Fuentes \textit{et al.} \cite[Eq. (16)]{fuentes89}.

\subsection*{Membrane deformation}

The force exerted on the particle induces a fluid motion which creates imbalance in the stress tensor across the membrane.
As a result, the membrane deforms elastically. We now compute the capsule deformation field resulting from a nearby point-force. To leading order in deformation, the displacement of the membrane is related to the fluid velocity via the no-slip relation given by Eq.~\eqref{no-slip-equation}.
From Eqs.~\eqref{vInside_r} and \eqref{vInside_t} we obtain
\begin{align}
  u_r &= \frac{1}{8\pi\eta i\omega} \infSumOne \left[ \frac{n+1}{2} a_n + b_n - c_{n+1} \right] \psi_{n-1} \, , \label{uRad} \\
  \bPi \vect{u} &= \frac{1}{8\pi\eta i\omega} \bigg[ \infSumOne \bigg( \frac{c_{n+3}}{n+2} -\frac{c_{n+1}}{n} + \frac{b_n}{n} \notag \\ 
  &+ \frac{n+3}{2n} a_n \bigg) \bPsi_{n-1} - \infSum \frac{n+1}{n+2} c_{n+3} \, \bGamma_n \bigg] \, , \label{uTang}
\end{align}
{where $\bPi$ denotes the projection operator defined as
\begin{equation}
 \bPi := \boldsymbol{1} - \eR \eR \, , \notag
\end{equation}
and
\begin{equation}
 \bGamma_n := \bPi \bgamma_n \, , \quad\quad \bPsi_n := \bPi \bNabla \psi_n \, . \notag
\end{equation}}

We define the frequency-dependent reaction tensor $R_{ij}$ relating the membrane displacement to the point-force as \cite{bickel07}
\begin{equation}
 u_i (\phi, \theta, \omega) = R_{ij} (\phi,\theta,\omega) F_j (\omega) \, . \notag
\end{equation}

In particular, by considering a harmonic driving force $F_i(t)=K_i e^{i\omega_0 t} $ of frequency $\omega_0$, which in the frequency domain has the form $F_i(\omega) = 2\pi K_i \delta (\omega - \omega_0)$, the membrane displacement in the temporal domain obtained upon inverse Fourier transform is calculated as
\begin{equation}
 u_i(\phi, \theta, t) = R_{ij} (\phi, \theta, \omega_0) K_j e^{i\omega_0 t} \, . \notag
\end{equation}

Explicit expression for the reaction tensor can readily be obtained from Eqs.~\eqref{uRad} and \eqref{uTang} upon separating out the force $\vect{F}$ in $\psi_{n-1}$, $\bPsi_{n-1}$ and $\bGamma_n$.

\begin{figure}[t]
 \includegraphics[scale=1]{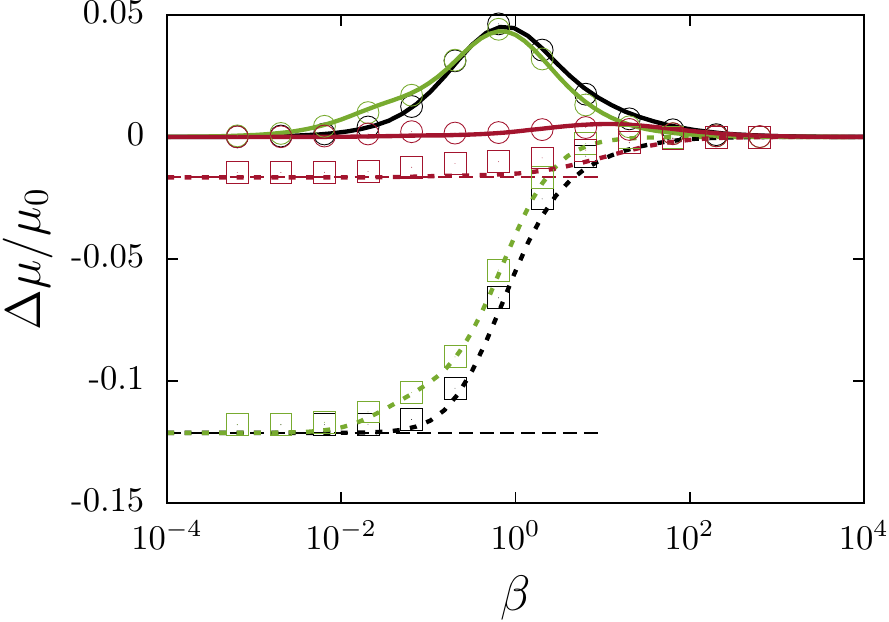}
 \caption{(Color online) Scaled frequency-dependent particle self-mobility correction versus the scaled frequency $\beta$ nearby a membrane endowed with only shearing (green), only bending (red) and both rigidities (black).
 The small particle of radius $b=1/10$ is at a distance $h=2b$.
 Here we take a Skalak ratio $C=1$ and a reduced bending modulus $\EB=2/3$.
 The theoretical predictions are presented as dashed lines for the real parts and as solid lines for the imaginary parts.
 Symbols refer to boundary integral simulations where the real and imaginary parts are shown as squares and circles, respectively.
 The horizontal dashed lines are the vanishing frequency limits predicted by Eqs.~\eqref{hardSphere} and \eqref{bendingLimit_Approximation}.
 }
 \label{anaNum_para}
\end{figure}

\begin{figure}[t]
 \includegraphics[scale=1]{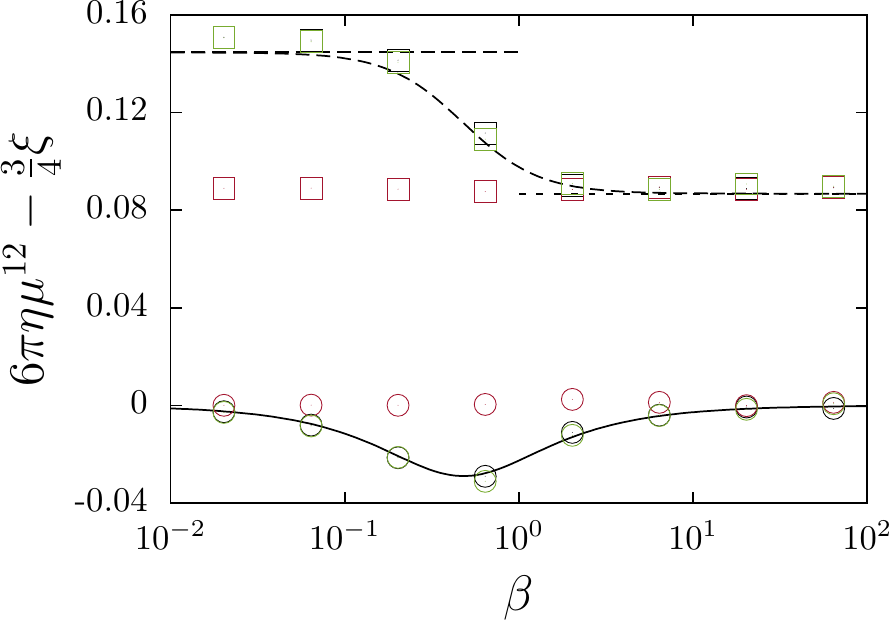}
 \caption{ (Color online) The scaled pair-mobility correction versus the scaled frequency nearby a membrane possessing only shearing (green), only bending (red) and both rigidities (black). 
 The analytical prediction given by Eq.~\eqref{mobilityCorrection_Pair} is shown as dashed line for the real part and as solid line for the imaginary part.
 Simulation results are shown as squares and circles for the real and imaginary parts, respectively.
 The horizontal long-dashed line is the hard-sphere limit predicted by Eq.~\eqref{pairMobi_hardSphere} where the short-dashed line is the viscous drop limit given by Eq.~\eqref{pairMobi_viscousDrop}.
 }
 \label{capsPairMobiPara}
\end{figure}

\section{Comparison with boundary integral simulations}\label{sec:comparisonWithBIM}

In order to assess the validity and accuracy of the point-particle approximation used throughout this work, we compare our analytical predictions with fully resolved numerical simulations.
The simulation method is based on the completed double layer boundary integral equation method (CDLBIEM)  \cite{phan92b, kohr04, zhao11, zhao12}, which has been proven to be perfectly suited for simulating deformable soft objects and solid particles in the vanishing Reynolds number regime.
Further technical details regarding the algorithm and its numerical implementation have been reported by some of us elsewhere, see e.g. Ref.~\onlinecite{daddi16b} and \onlinecite{guckenberger16}.

In Fig.~\ref{anaNum_para}, we show the scaled particle self-mobility correction versus the scaled frequency predicted theoretically by Eq.~\eqref{mobilityCorrection}.
The solid particle of radius $b=1/10$ is positioned at $h=2b$ close to a large spherical capsule.
Here we take the same simulation parameters as in part~I for a Skalak ratio $C=1$ $(\lambda=2)$ and a reduced bending modulus $\EB = 2/3$.
We also show results for an idealized membrane with pure shearing (green) and pure bending (red).
 
We observe that shearing resistance manifests itself in a more pronounced way compared to bending.
The mobility correction nearby a hard-sphere is recovered only if the membrane possesses a non-vanishing resistance towards shearing, in line with theoretical predictions.
A very good agreement is obtained between analytical predictions and boundary integral simulations over the whole range of applied frequencies.

Next, we turn our attention to the motion of the capsule induced by the nearby solid particle.
In Fig.~\ref{capsPairMobiPara} we plot the scaled pair-mobility correction versus the scaled frequency as predicted theoretically by Eq.~\eqref{mobilityCorrection_Pair}.
We observe that the correction for a membrane with pure shearing is almost identical to that of a membrane with both shearing and bending resistances, thus confirming the dominant contribution of shearing to the pair-mobility.
For small forcing frequencies, the correction approaches that near a hard-sphere given by Eq.~\eqref{pairMobi_hardSphere}.
On the other hand, the correction approaches that near a viscous drop for high frequencies as given by Eq.~\eqref{pairMobi_viscousDrop}.
The correction nearby a membrane with pure bending remains typically  constant upon changing the actuation frequency, and equals to that predicted nearby a viscous drop, in agreement with theoretical predictions.

\begin{figure}[!ht]
 \includegraphics[scale=1]{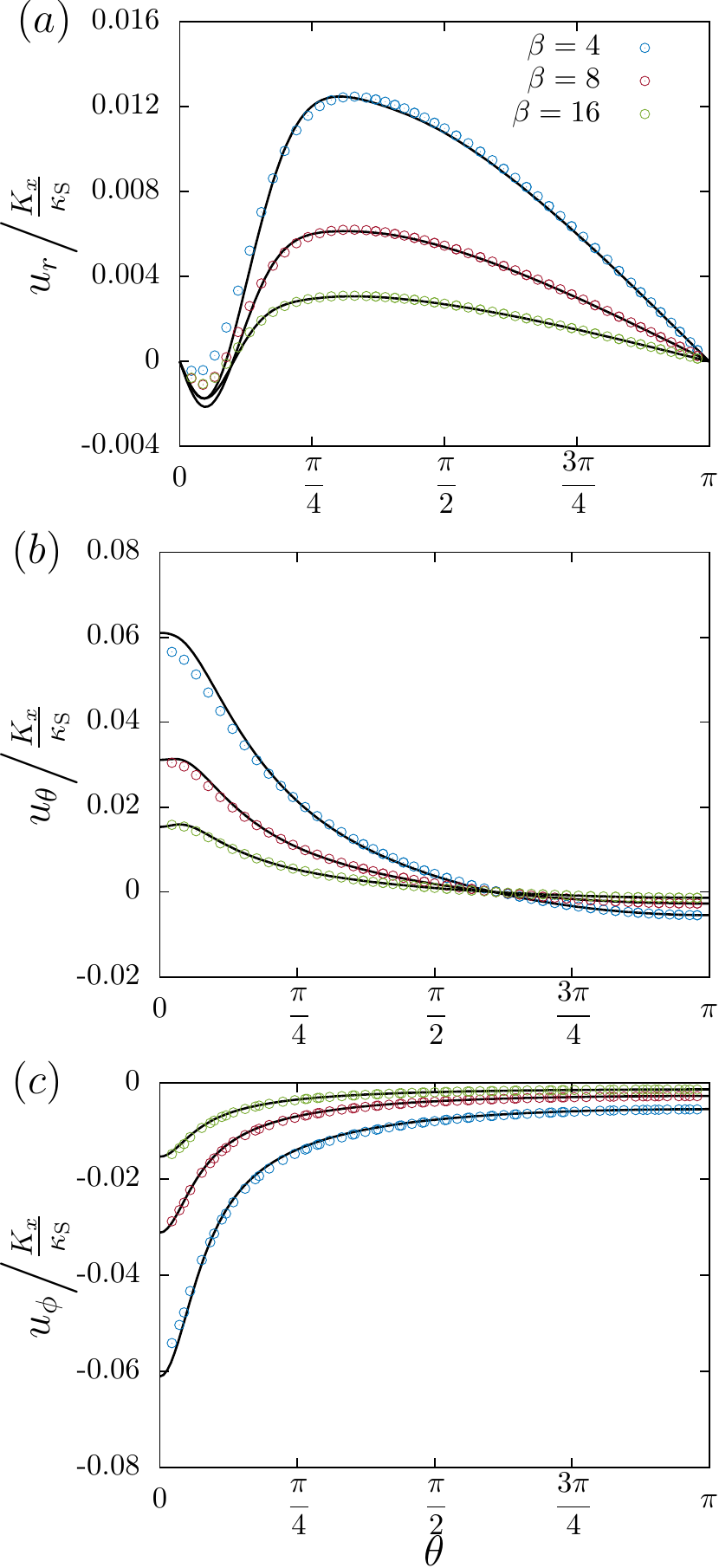}
 \caption{(Color online) The membrane displacement versus the polar angle $\theta$ in the plane of maximum displacement (the plane $\phi=0$ for $u_r$ and $u_\theta$ and the plane $\phi=\pi/2$ for $u_\phi$) for three scaled forcing frequencies $\beta$ at quarter period for $t\omega_0 = \pi/2$.
 Solid lines are the theoretical predictions obtained from Eqs.~\eqref{uRad} and \eqref{uTang} and  symbols are boundary integral simulations.
 }
 \label{deformationParallel}
\end{figure}

In Fig.~\ref{deformationParallel}, we present a comparison between analytical prediction and boundary integral simulations of the capsule deformation for a membrane possessing both shearing and bending resistances, using the same parameters as in Fig.~\ref{anaNum_para}.
The displacement field is shown in the plane of maximal deformation (the plane $\phi=0$ for $u_r$ and $u_\theta$ and the plane $\phi=\pi/2$ for $u_\phi$), plotted at quarter period for which $t \omega_0 = \pi/2$ ~i.e.\@ when the oscillating particle reaches its maximal position.
We observe that the radial displacement vanishes at the capsule poles and shows a non-monotonic dependence on the polar angle $\theta$.
On the other hand, the in-plane displacements $u_\theta$ and $u_\phi$ are monotonically decreasing functions of $\theta$ and reach their maximum at $\theta=0$.
We observe that the in-plane displacement along the membrane is about five times larger than the radial displacement, in contrast to the axisymmetric motion where the radial displacement is found to be about three times larger than tangential displacement.
By analyzing the displacement at various actuation frequencies, we observe that larger frequencies induce smaller deformation as the capsule membrane does not have enough time to respond to the fast wiggling particle.
As shown in part~I for typical situations, taking a forcing frequency $\beta=4$ induces a maximal membrane deformation of about 1~\% of its initially undeformed radius.
As a result, departure from sphericity is negligible and the system can accurately be studied within the frame of the linear theory of elasticity adopted throughout this work.
The analytical predictions are found to be in a very good agreement with boundary integral simulations.

\section{Conclusions}\label{sec:conclusions}

This work, together with an earlier paper\cite{daddi17b}, provides a complete solution of the hydrodynamic problem of flow induced by a point-force acting close to an elastic spherical capsule. The answer is formulated in terms of the Green's function. The problem for the force acting along the symmetry axis of the system has been treated in the first part of our considerations, while here we have extended the results to account for the force being tangential to the surface of the sphere. 
Together with the result of part~I, the fluid flow field and thus the particle mobility functions can then be obtained for an arbitrary direction of motion.
The solution has been found using the image technique. Giving all the technical details, we have done our calculations for the two cases of a membrane exhibiting resistance against shearing and bending, respectively, and explicit formulas have been presented. The same technique has been used to assess the combined effect of the two deformation modes.

We have then used the solution to characterize various dynamic effects related to this motion. To explore the effect of confinement on the motion of the particle, we have calculated the leading-order  frequency-dependent hydrodynamic self-mobility of a small solid sphere moving close to the capsule.  
We have shown that shearing resistance induces a second low-frequency peak resulting from the membrane normal displacement.
Moreover, we have demonstrated, in agreement with previous studies in different complex geometries, that in the vanishing frequency limit the particle self-mobility near a hard sphere is recovered only when the membrane possesses a  non-zero resistance against shearing. 
{By applying the fluctuation-dissipation theorem, we find that the elastic nature of the membrane introduces a memory in the system resulting to a long-lived subdiffusive regime on nearby Brownian particles.
The planar membrane assumption is found to be not valid for strongly curved membranes where the steady excess MSD is significantly smaller than that predicted for the planar case.
}

The effect of the point-force on the capsule has been quantified in two ways. Firstly, we have calculated and analyzed the pair-mobility function, which is determined solely by the shearing properties of the membrane. We have shown it to be well described by a Debye-like model with a single relaxation time. Secondly, we have computed, in leading order, the deformation of the membrane due to the action of a point-force nearby. 

All the theoretical results shown in the paper have been favorably verified in representative cases by fully resolved numerical simulations for a truly extended particle using the completed double layer boundary integral method.

\acknowledgments
ADMI and SG thank the Volkswagen Foundation for financial support and acknowledge the Gauss Center for Supercomputing e.V. for providing computing time on the GCS Supercomputer SuperMUC at Leibniz Supercomputing Center. 
This work has been supported by the Ministry of Science and Higher Education of Poland via the Mobility Plus Fellowship awarded to ML.
This article is based upon work from COST Action MP1305, supported by COST (European Cooperation in Science and Technology).

\appendix

\section{Membrane mechanics}\label{appendix:membraneMechanics}

In this appendix, we derive the traction jumps across a membrane endowed with shearing and bending rigidities expressed in the spherical coordinates system for an asymmetric deformation.
Here we follow the convention in which the symbols for the radial, azimuthal and polar angle coordinates are taken as $r$, $\phi$ and $\theta$ respectively,
with the corresponding orthonormal basis vectors $\eR$, $\ePhi$ and $\eThe$.

Similarly, all the lengths shall be scaled by the capsule radius $a$.
We denote by $\vect{a} = \eR$ the position vector of the points situated at the undisplaced membrane.
After deformation, the vector position reads
\begin{equation}
 \vect{r} = (1+u_r)\eR + u_\theta \eThe + u_\phi \ePhi \, ,
\end{equation}
where $\vect{u}$ denotes the displacement vector field.
In the following, capital roman letters will be reserved for the undeformed state while small letters for the deformed.
The spherical membrane can be defined by the covariant base vectors $\gOne := \vect{r}_{,\theta}$ and $\gTwo := \vect{r}_{,\phi}$, where commas in indices denote spatial derivatives.
The unit vector $\vect{n}$ normal to the membrane is defined in such a way to form a direct trihedron with $\gOne$ and $\gTwo$.
The covariant base vectors are
\begin{subequations}
 \begin{align}
  \gOne =&  (u_{r,\theta} - u_{\theta}) \eR + (1 + u_r + u_{\theta,\theta} ) \eThe + u_{\phi, \theta} \ePhi \, , \\
  \gTwo =&  (u_{r,\phi} - u_{\phi} \sin \theta) \eR + (u_{\theta,\phi} - u_{\phi} \cos \theta) \eThe \\ \nonumber
        &+ \left( (1 + u_r) \sin \theta + u_{\theta} \cos \theta + u_{\phi,\phi} \right) \ePhi \, ,
 \end{align}
\end{subequations}
and the linearized unit normal vector reads
\begin{equation}
 \vect{n} = \eR - \left( u_{r,\theta} - u_{\theta} \right) \eThe - \left( \frac{u_{r,\phi}}{\sin\theta} - u_{\phi} \right) \ePhi \, . 
\end{equation}

Note that $\gOne$ and $\gTwo$ have (scaled) length dimension while the normal vector $\vect{n}$ is dimensionless.
In the deformed state, the components of the metric tensor are defined by the scalar product $g_{\alpha\beta} = \vect{g}_{\alpha} \cdot \vect{g}_{\beta}$.
The contravariant tensor $g^{\alpha\beta}$, being the inverse of the metric tensor is linearized as
\begin{equation}
 g^{\alpha\beta} = \left(
                   \begin{array}{cc}
                    1 - 2\epsilon_{\theta\theta} & -\frac{2\epsilon_{\theta\phi}}{\sin\theta} \\
                    -\frac{2\epsilon_{\theta\phi}}{\sin\theta} & \frac{1-2\epsilon_{\phi\phi}}{\sin^2 \theta}
                   \end{array}
		   \right) \, ,
		    \label{contravariantTensor}
\end{equation}
wherein $\epsilon_{\alpha\beta}$ are the components of the in-plane strain tensor expressed in spherical coordinates as \cite{sadd09}
\begin{subequations}
 \begin{align}
  \epsilon_{\theta\theta} &=  (u_r + u_{\theta,\theta}) \, , \\
  \epsilon_{\theta\phi} &= \frac{1}{2} \left( \frac{u_{\theta,\phi}}{\sin\theta}  + u_{\phi,\theta} - u_{\phi} \cot \theta \right) \, , \\ 
  \epsilon_{\phi\phi} &= \left( u_r + \frac{u_{\phi,\phi}}{\sin\theta} + u_{\theta} \cot \theta \right) \, .
 \end{align}
\end{subequations}

The contravariant tensor in the undeformed state $G^{\alpha\beta}$ is readily obtained by considering a vanishing strain tensor in Eq.~\eqref{contravariantTensor}.

\subsection{Shearing contribution}

In this subsection, we derive the traction jump equations across a membrane endowed with a pure shearing resistance.
The two strain tensor invariants are given by Green and Adkins as \cite{green60, zhu14, zhu15}
\begin{subequations}
 \begin{align}
  I_1 &= G^{\alpha\beta} g_{\alpha\beta} - 2 \, , \\
  I_2 &= \det G^{\alpha\beta} \det g_{\alpha\beta} - 1 \, .
 \end{align}
\end{subequations}

The stress tensor contravariant components $\tau^{\alpha\beta}$ can be obtained provided knowledge of the constitutive elastic law of the membrane, whose areal strain energy functional is $W (I_1,I_2)$, such that \cite{lac04}
\begin{equation}
 \tau^{\alpha\beta} = \frac{2}{\JS} \frac{\partial W}{\partial I_1} \, G^{\alpha\beta} + 2\JS \frac{\partial W}{\partial I_2} \, g^{\alpha\beta} \, ,
 \label{stressTensor}
\end{equation}
wherein $\JS := \sqrt{1+I_2}$ is the Jacobian determinant, quantifying the ratio between deformed and undeformed local areas.
In the linear theory of elasticity, $\JS \simeq 1 + e$, with $e := \epsilon_{\theta\theta} + \epsilon_{\phi\phi}$ being the trace of the in-plane strain tensor, also know as the dilatation function \cite{love13}.
In this work, we use the Skalak model to describe the elastic properties of the capsule membrane such that \cite{sui08PRE, krueger12, gekle16, baecher17}
\begin{equation}
 W(I_1, I_2) = \frac{\kS}{12} \left( I_1^2 + 2I_1-2I_2 + C I_2^2 \right) \, ,
 \label{skalakEquation}
\end{equation}
where $C:=\kA/\kS$ is a dimensionless parameter defined as the ratio between the area expansion and shear modulus.
We note that for $C=1$, the Skalak model and the Neo-Hookean model are equivalent for small deformations \cite{lac04}.
Upon plugging Eq.~\eqref{skalakEquation} into Eq.~\eqref{stressTensor}, the linearized in-plane stress tensor reads
\begin{equation}
 \tau^{\alpha\beta} = \frac{2 \kS}{3}
 \left(
 \begin{array}{cc}
  \epsilon_{\theta\theta} + Ce & \frac{\epsilon_{\theta\phi}}{\sin\theta} \\
  \frac{\epsilon_{\theta\phi}}{\sin\theta} & \frac{\epsilon_{\phi\phi} + Ce}{\sin^2 \theta} 
 \end{array}
 \right) \, .
\end{equation}

The membrane equilibrium equations which balance the elastic and external forces read
\begin{subequations}
 \begin{align}
  \nabla_{\alpha} \tau^{\alpha\beta} + \Delta f^{\beta} &= 0 \, , \label{Equilibrium_Tangential} \\
  \tau^{\alpha\beta} b_{\alpha\beta} + \Delta f^{n} &= 0 \, , \label{Equilibrium_Normal}
 \end{align}
\end{subequations}
where $\Delta \vect{f} = \Delta f^{\beta} \vect{g}_{\beta} + \Delta f^{n} \vect{n} $ is the traction jump and $\nabla_{\alpha}$ stands for the covariant derivative defined for a second-rank tensor as \cite{deserno15}
\begin{equation}
 \nabla_{\alpha} \tau^{\alpha\beta} = \tau^{\alpha\beta}_{,\alpha} + \Gamma_{\alpha\eta}^{\alpha} \tau^{\eta\beta} + \Gamma_{\alpha\eta}^{\beta} \tau^{\alpha\eta} \, ,
\end{equation}
and $\Gamma_{\alpha\beta}^{{\lambda}}$ are the Christoffel symbols of the second kind defined as \cite{synge69} 
\begin{equation}
 \Gamma_{\alpha\beta}^{{\lambda}} = \frac{1}{2} g^{{\lambda}\eta} \left( g_{\alpha\eta,\beta} + g_{\eta\beta,\alpha} - g_{\alpha\beta, \eta} \right) \, .
\end{equation}

Further, $b_{\alpha\beta}$ is the second fundamental form (curvature tensor) defined as
\begin{equation}
 b_{\alpha\beta} = \vect{g}_{\alpha,\beta} \cdot \vect{n} \, .
\end{equation}

In spherical coordinates, the non-vanishing components of the Christoffel symbols at zeroth order are $\Gamma_{\phi\theta}^{\phi} = \Gamma_{\theta\phi}^{\phi} = \cot \theta$ 
and $\Gamma_{\phi\phi}^{\theta} = -\sin\theta\cos\theta$.
We find after some algebra that the tangential traction jumps across the membrane as given by Eq.~\eqref{Equilibrium_Tangential} read
\begin{subequations}
 \begin{align}
   \tau^{\theta\theta}_{,\theta} +  \tau^{\theta\phi}_{,\phi} + \Gamma_{\phi\theta}^{\phi} \tau^{\theta\theta} + \Gamma_{\phi\phi}^{\theta} \tau^{\phi\phi} + \Delta f^{\theta} &= 0 \, , \\
   \tau^{\theta\phi}_{,\theta} +  \tau^{\phi\phi}_{,\phi} + \left( 2\Gamma_{\phi\theta}^{\phi} + \Gamma_{\theta\phi}^{\phi} \right) \tau^{\theta\phi} + \Delta f^{\phi} &= 0 \, .
 \end{align}
\end{subequations}

At zeroth order, the non-vanishing components of the curvature tensor are $b_{\theta\theta} = -1$ and $b_{\phi\phi} = -\sin^2\theta$.
For the normal traction jump Eq.~\eqref{Equilibrium_Normal} we obtain
\begin{equation}
 -\tau^{\theta\theta} - \sin^2 \theta \tau^{\phi\phi} + \Delta f^{n} = 0 \, .
\end{equation}

After substitution and writing the projected equations in the spherical coordinates basis, we immediately get the following set of equations for the traction jump,
\begin{subequations}
 \begin{align}
   \Delta f_{\theta} &= -\frac{2\kS}{3} \bigg( {(1+C)}\epsilon_{\theta\theta,\theta} + C\epsilon_{\phi\phi,\theta} \notag \\
                     &+ \frac{\epsilon_{\theta\phi,\phi}}{\sin\theta} 
		+ (\epsilon_{\theta\theta} - \epsilon_{\phi\phi}) \cot \theta \bigg)  \, ,  \\
   \Delta f_{\phi} &= -\frac{2\kS}{3}\bigg( \epsilon_{\theta\phi,\theta} + \frac{1}{\sin\theta} \left( C \epsilon_{\theta\theta,\phi} + {(1+C)} \epsilon_{\phi\phi,\phi} \right) \notag \\
		  &+ 2\epsilon_{\theta\phi} \cot \theta \bigg)  \, , \label{equilibriumEqn_Theta} \\
   \Delta f_n &= \frac{2\kS}{3} (1+2C) \left( \epsilon_{\theta\theta} + \epsilon_{\phi\phi} \right) \, .
 \end{align}
 \label{equilibriumEqn}
\end{subequations}

It is worth to mention here that for curved membranes, the normal traction jump does not vanish in the \emph{plane stress} formulation employed throughout this work as the zeroth order in the curvature tensor is not identically null.
In fact, this is not the case for a planar elastic membrane where the resistance to shearing introduces a jump in the tangential traction jumps only \cite{daddi16, daddi16b, daddi16c}.
By substituting $\epsilon_{\theta\theta}$, $\epsilon_{\phi\phi}$ and $\epsilon_{\theta\phi}$ with their expressions, Eqs.~\eqref{equilibriumEqn} turn into the traction equations given by Eq.~\eqref{jump_Shear_main} of the main text.
In the following, the traction jump equations across a membrane with a bending rigidity shall be derived.


\subsection{Bending contribution}

For the membrane resistance towards bending, we use the linear isotropic model, in which the bending moment is related to the curvature tensor via \cite{pozrikidis01}
\begin{equation}
 M_{\alpha}^{\beta} = -\kB \left( b_{\alpha}^{\beta} - B_{\alpha}^{\beta} \right) \, ,
\end{equation}
where $\kB$ is the membrane bending modulus.
This model is equivalent to the Helfrich model for small deformations \cite{Guckenberger_preprint}.
The mixed version of the curvature tensor~$b_{\alpha}^{\beta}$ is related to its covariant representation by $b_{\alpha}^{\beta} = b_{\alpha\delta} g^{\delta\beta}$.
The contravariant components of the transverse shearing vector $\vect{Q}$ can be obtained from a local torque balance with the applied moment as $Q^{\beta} = \nabla_{\alpha} M^{\alpha\beta}$.
We note that the raising and lowering indices operations implies that $M^{\alpha\beta} = M_\delta^\alpha g^{\delta \beta}$.
Therefore, the components of the shearing force read
\begin{subequations}
 \begin{align}
  Q^{\theta} &= -{\kB} \Big[ \left(1-\cot^2\theta\right)u_{r,\theta} + u_{r,\theta\theta} \cot\theta   \\ \nonumber
           &+ u_{r,\theta\theta\theta} +(1+\cot^2\theta) \left( u_{r,\phi\phi\theta} - 2u_{r,\phi\phi} \cot\theta \right)  \Big]  \, , \\
  Q^{\phi} &= -{\kB} (1+\cot^2\theta) \Big(  u_{r,\phi\theta} \cot\theta + 2u_{r,\phi} \\ \nonumber
  &+ u_{r,\phi\theta\theta}   + (1+\cot^2\theta) u_{r,\phi\phi\phi}  \Big) \, .
 \end{align}
\end{subequations}

The equilibrium equations read
\begin{subequations}
 \begin{align}
  -b_{\alpha}^{\beta} Q^{\alpha} + \Delta f^{\beta} &= 0 \, , \\
  \nabla_{\alpha} Q^{\alpha} + \Delta f^{n} &= 0 \, ,
 \end{align}
\end{subequations}
where for a first-rank tensor the covariant derivative is defined as $\nabla_{\beta} Q^{\alpha} = \partial_{\beta} Q^{\alpha} + \Gamma_{\beta\delta}^{\alpha} Q^{\delta}$.
By substituting $Q^\theta$ and $Q^\phi$ with their expressions, we thus obtain the traction jumps given by Eqs.~\eqref{jump_Bending_main} of the main text.


\section{Determination of the unknown coefficients}\label{appendix:determination}

{In this appendix, we present technical details regarding the determination of the unknown coefficients ($A_n$, $B_n$, $C_n$ for the reflected flow, and $a_n$, $b_n$, $c_n$ for the inner flow).
For that purpose, we first project the velocities on the surface of the membrane onto the radial and tangential directions following the approach of Fuentes {\it et al} \cite{fuentes88,fuentes89}.}

\subsection{Velocity projections}

For the radial projection, we use the following identities 
\begin{subequations}
\begin{align}
  \eR\cdot\Nabla\psi_{n-1} &= -\frac{n+1}{r} \, \psi_{n-1}\, , \label{radialProjectionRelation_1} \\ 
  \eR\cdot\R\psi_{n-1} &= r \psi_{n-1} \, , \\
  \eR\cdot\bgamma_{n-1} &= -\frac{1}{r} \, \psi_{n-2} \, , \\
  \eR\cdot(\T\times\R)\, \vp_{n-1} &= 0 \, .
\end{align}
\end{subequations}
Moreover, the projection of Eq.~\eqref{eliminateFPhi} onto the radial direction leads to
\begin{equation}
 \eR \cdot \F \vp_{n} = \frac{1}{2n+1} \left( \frac{\psi_{n-2}}{r} - r \psi_n \right) \, . \label{radialProjectionRelation_5}
\end{equation}
Therefore, the radial components can all be expressed in terms of a single harmonic $\psi_n$. 
Using these identities in Eqs.~\eqref{Stokeslet_finalize}, \eqref{imageSystem_finalized} and \eqref{insideSolution_finalized}, we obtain
\begin{align} 
\label{vStok_r}
 8\pi\eta \vStokcom_r &= \infSumOne \left[ \frac{n-2}{2n-1} \frac{r^{2n}}{R^{n}} - \frac{n}{2n+3} \frac{r^{2n+2}}{R^{n+2}} \right] \psi_{n-1} \, ,  \\  \label{vIm_r}
 8\pi\eta v_r^* &= \infSumOne \left[ -\frac{n+1}{2n-1}r A_{n-1} +\frac{n+3}{2n+3} \frac{A_{n+1}}{r}\right. \\ \nonumber
 &+\left. 2(n+1)\frac{B_{n-1}}{r} -\frac{C_n}{r} \right] \psi_{n-1} \, ,  \\ \label{vInside_r}
 8\pi\eta v_r^{(i)} &= \infSumOne \bigg[ \frac{n+1}{2} a_n r^{2n+2} \\ \nonumber 
 &+ b_n r^{2n}- c_{n+1} r^{2n} \bigg] \psi_{n-1} \, . 
\end{align}

For the projection onto the tangential direction, we need to use the orthogonality properties of spherical harmonics on a spherical surface. To this end, we introduce the following notation for the average of a given scalar quantity $M$ over a sphere,
\begin{equation}
\avg{M} := \frac{1}{2\pi} \int_0^{2\pi} \int_0^\pi M \sin \theta \, \Intd \theta \, \Intd \phi \, , \label{averagingDefinition}
\end{equation}
which we will use extensively for writing the orthogonality properties of the considered functions. In particular, we have
\begin{align}
 \avg{\varphi_{m-1} \varphi_{n-1} } &= \frac{2}{2n+1} \frac{\delta_{mn}}{r^{2n+2}} \, , \notag \\
 \avg{\psi_{m-1} \psi_{n-1} }   &= \frac{n(n+1)}{2n+1} \frac{\delta_{mn}}{r^{2n+2}} \, . \notag
\end{align}
We also define the operator 
\begin{equation}
 \bPi := \boldsymbol{1} - \eR \eR \, , \notag
\end{equation}
which projects vectors on a plane tangent to the spherical membrane surface. By applying the projection operator to Eq.~\eqref{eliminateFPhi}, we obtain
\begin{equation}
 (n+1) (\bPi \F) \vp_n = \frac{1}{2n+1} \left( \bPsi_{n-2} - r^2 \bPsi_n \right)-\bGamma_{n-1}, \label{tangentialProjectionRelation_1}
\end{equation}
where we have defined 
\begin{equation}
 \bGamma_n := \bPi \bgamma_n \, , \quad\quad \bPsi_n := \bPi \bNabla \psi_n \, . \notag
\end{equation}
We also note the relation
\begin{equation}
 \begin{split}
  (2n &- 1)\bPi  (\T\times\R)\vp_{n-1} \\
       &= \bGamma_{n-3}-r^2 \bGamma_{n-1} + (2n-3) (\bPi \F) \vp_{n-2} \, ,  \label{secondProjected}
 \end{split}
\end{equation}
which upon using Eq.~\eqref{tangentialProjectionRelation_1} gives
\begin{equation}
 \begin{split}
  \label{tangentialProjectionRelation_2}
     (2n-1)\bPi (\T\times\R)\vp_{n-1} &= \frac{1}{n-1} \left( \bPsi_{n-4} - r^2 \bPsi_{n-2} \right) \\ 
     &-\frac{n-2}{n-1} \bGamma_{n-3} - r^2 \, \bGamma_{n-1} \, . 
 \end{split}
\end{equation}

Applying the projection relations Eq.~\eqref{tangentialProjectionRelation_1} and \eqref{tangentialProjectionRelation_2} to Eqs.~\eqref{Stokeslet_finalize}, \eqref{imageSystem_finalized} and \eqref{insideSolution_finalized}, we finally obtain
\begin{widetext}
\begin{align}
   8\pi\eta \, \bPi \vStok &= \infSumOne \left[\frac{n-2}{(2n-1){n}}\frac{{r^{2n+1}}}{R^n} - \frac{n}{(n+2)(2n+3)}\frac{{r^{2n+3}}}{R^{n+2}}\right] \bPsi_{n-1}
   + \infSum  -\frac{2}{n+2} \frac{{r^{2n+3}}}{R^{n+2}} \, \bGamma_{n} \, , \label{vStok_t} \\
   8\pi\eta \, \bPi \vect{v}^* &= \infSumOne \left[ -\frac{n}{(n+2)(2n+3)} A_{n+1} + \frac{n-2}{n(2n-1)}\, r^2 A_{n-1} - 2B_{n-1} \right] \bPsi_{n-1}
                      + \sum_{{n=0}}^\infty \left[ C_n - \frac{2}{n+2} A_{n+1} \right] \bGamma_n \, , \label{vIm_t} \\
   8\pi\eta \, \bPi \vect{v}^{(i)} &= \infSumOne \left[ \frac{r^{2n+3}}{n+2} c_{n+3}-\frac{r^{2n+1}}{n} c_{n+1}+b_n \frac{r^{2n+1}}{n}+a_n \frac{n+3}{2n} r^{2n+3} \right] \bPsi_{n-1} + \infSum -\frac{n+1}{n+2} \, r^{2n+3} c_{n+3} \, \bGamma_n \, . \label{vInside_t}
\end{align}
\end{widetext}

The functions $\bPsi_{n-1}$ and $\bGamma_n$ satisfy the following orthogonality relations
\begin{align}
 \avg{ \bPsi_{m-1} \cdot \bPsi_{n-1}  } &= \frac{n^2(n+1)^2}{2n+1} \frac{\delta_{mn}}{r^{2n+4}} \, , \label{Psi_DOT_Psi} \\
 \avg{ \bGamma_m \cdot \bGamma_n  } &=\frac{4(n+1)^3}{(2n+1)(2n+3)} \frac{\delta_{mn}}{r^{2n+4}}\, , \label{Gamma_DOT_Gamma}
\end{align}
and also for cross terms
\begin{equation}
 \avg{ \bPsi_{m-1} \cdot \bGamma_n } = \frac{n^2(n+1)}{2n+1} \frac{\delta_{mn}}{r^{2n+4}} \, . \label{Psi_DOT_Gamma}
\end{equation}
We note that their derivatives with respect to $r$ needed for the computation of stresses can be obtained from
\begin{align}
 {\bPsi_{n-1,r}} &= - \frac{n+2}{r} \, \bPsi_{n-1} \, , \notag \\
 {\bGamma_{n,r}} &= - \frac{n+2}{r} \, \bGamma_n \, . \notag
\end{align}

Having introduced these tools, we now proceed to the calculation of the fluid velocity coefficients.

\subsection{Pressure field}

The pressure field can be found by multipole expansion. 
The general form of the pressure $p$ in the exterior fluid is written as a sum of exterior and interior harmonics as
\begin{equation}
 8\pi p = \infSumOne \left( S_n + Q_n r^{2n+1} \right) \psi_{n-1} \, . \notag
\end{equation}
The coefficients $S_n$ and $Q_n$ can be related to the coefficients of the velocity thanks to the Stokes equation~\eqref{Stokes:Momentum}, leading to
\begin{equation}
 S_n = -2A_{n-1} \, , \qquad Q_n = -\frac{2}{R^{n+2}} \, . \notag
\end{equation}

For the fluid inside the capsule, all harmonics of negative order that lead to a singularity at the origin should be discarded, thus reducing the form of the pressure to
\begin{equation}
 8\pi p^{(i)} = \infSumOne p_n r^{2n+1} \psi_{n-1} \, ,  \notag
\end{equation}
leading upon using Eq.~\eqref{Stokes:Momentum_Inside} to
\begin{equation}
 p_n = \frac{(n+1)(2n+3)}{n} \, a_n \, . \notag
\end{equation}

\subsection{Continuity of velocity}

After substituting the radially projected velocities given by Eqs.~\eqref{vStok_r} through \eqref{vInside_r} into Eq.~\eqref{BC:v_r}, the continuity of the radial component at the membrane leads to
\begin{align} 
  \label{continuity_1} 
  &\frac{n+3}{2n+3} A_{n+1}- \frac{n+1}{2n-1} A_{n-1} + 2(n+1) B_{n-1}  \\ \nonumber
  -&C_n + \frac{n-2}{2n-1} \frac{1}{R^n}-\frac{n}{2n+3} \frac{1}{R^{n+2}} \\ \nonumber
  =& \frac{n+1}{2} a_n+b_n-c_{n+1},
\end{align}
in direct analogy with Fuentes {\it et al}\cite{fuentes89}. 

Substituting Eqs.~\eqref{vStok_t} through \eqref{vInside_t} into Eq.~\eqref{BC:v_phi} and \eqref{BC:v_the}, the continuity of the tangential velocity across the membrane leads to the two following equations
\begin{align} \label{continuity_2}
 &-\frac{n}{(n+2)(2n+3)}A_{n+1}+\frac{n-2}{n(2n-1)}A_{n-1}  \\ \nonumber
 &{-2B_{n-1}}+\frac{n-2}{n(2n-1)}\frac{1}{R^n}-\frac{n}{(n+2)(2n+3)}\frac{1}{R^{n+2}}  \\ \nonumber
   &= \frac{c_{n+3}}{n+2}-\frac{c_{n+1}}{n}+\frac{b_n}{n}+\frac{n+3}{2n}a_n \, ,  \\ \label{continuity_3}
 &\frac{2}{n+2} A_{n+1}-	C_n+\frac{2}{n+2} \frac{1}{R^{n+2}} = \frac{n+1}{n+2} c_{n+3} \, . 
\end{align}
We note that Fuentes \textit{et al.} \cite[p.~64]{fuentes89} reported ${-2B_{n-1}}$ with an erroneous plus sign, which we correct here.

Solving Eqs.~\eqref{continuity_1}, \eqref{continuity_2} and \eqref{continuity_3} for the unknown coefficients inside the capsule $a_n$, $b_n$ and $c_n$ leads to
\begin{align} \label{eq_1}
 a_n &= A_{n-1}-\frac{2n^2+7n+3}{2n^2+5n+3} A_{n+1}-2(2n+1)B_{n-1} \\ \nonumber
 &+\frac{2n+1}{n+1}C_n - \frac{2n}{2n^2+5n+3} \frac{1}{R^{n+2}} \, ,  \\ \label{eq_2}
 b_n &= -\frac{2n^3+n^2-10n+3}{2(n-1)(2n-1)} A_{n-1}+\frac{n+3}{2}A_{n+1} \\ \nonumber
 &+(2n^2+5n+3)B_{n-1}-\frac{n}{n-1} C_{n-2}  \\ \nonumber
     &-\frac{2n+3}{2} C_n+\frac{n(n+1)}{(2n-1)(n-1)} \frac{1}{R^n} \, ,  \\ \label{eq_3}
 c_n &= \frac{2}{n-2} A_{n-2} - \frac{n-1}{n-2} C_{n-3} + \frac{2}{n-2} \frac{1}{R^{n-1}} \, .
\end{align}

\subsection{Discontinuity of stress tensor}

Expressions for $A_n$, $B_n$ and $C_n$ can be determined from the discontinuity of the fluid stress tensor across the membrane.
In order to gauge the effect of membrane shearing and bending on the particle mobility, we hereafter consider shearing and bending effects separately.

\subsection*{Pure shearing}

For the sake of clarity, we write the radial and tangential velocities respectively stated by Eqs.~\eqref{vStok_r}-\eqref{vInside_r} and \eqref{vStok_t}-\eqref{vInside_t} as
\begin{equation}
  v_r           = \infSumOne \rho_n \psi_{n-1} \, , \qquad
 \bPi \vect{v} = \infSumOne \alpha_n \bPsi_{n-1} + \infSum \beta_n \bGamma_n \, ,  \notag
\end{equation}
for the fluid velocity outside the capsule wherein $\rho_n$, $\alpha_n$ and $\beta_n$ are functions of $r$ only.
Analogous expressions can be written for the radial and tangential velocities inside with the corresponding coefficients $\rho_n^{(i)}$, $\alpha_n^{(i)}$ and $\beta_n^{(i)}$.

Eqs.~\eqref{BC:sigma_r_phi} and \eqref{BC:sigma_r_the} with the shearing part only, as given by Eqs.~\eqref{jump_u_The_Shear_main} and \eqref{jump_u_Phi_Shear_main}, can be cast in the following form
\begin{equation}
 \begin{split}
  \label{tangentialTractionJump_Shear}
 \infSumOne \tilde{\alpha}_n \bPsi_{n-1} &+\infSum \tilde{\beta}_n \bGamma_n = \\ 
 & \infSumOne \alpha_n \vect{F}_n+\infSum \beta_n \vect{G}_n+\infSumOne \rho_n \vect{f}_n \, , 
 \end{split}
\end{equation}
where
\begin{equation}
 \tilde{\alpha}_n = \alpha_{n,r}-\alpha_{n,r}^{(i)}-(n+2) \left( \alpha_n-\alpha_n^{(i)} \right) \, ,  \notag
\end{equation}
and analogously for $\tilde{\beta}_n$.
Expressions for $\vect{F}_n$, $\vect{G}_n$ and $\vect{f}_n$ can readily be obtained by identification.
Multiplying both members of Eq.~\eqref{tangentialTractionJump_Shear} by $\bPsi_{m-1}$ and by $\bGamma_m $, and
averaging over the surface of the sphere allows us to use the following orthogonality relations  
\begin{align}
 \avg{ \vect{F}_n\cdot\bPsi_{m-1} } &= \frac{\alpha n^2(n+1)^2}{2n+1} \left( n(n+1){\lambda}-1 \right) \, , \notag \\
 \avg{ \vect{G}_n\cdot\bPsi_{m-1} } &= \frac{\alpha n^2(n+1)}{2n+1} \left( n(n+1){\lambda}-1 \right) \, , \notag \\
 \avg{ \vect{f}_n\cdot\bPsi_{m-1} } &= -\frac{\alpha n^2(n+1)^2}{2n+1} \left( 2{\lambda}-1 \right) \, , \notag \\
 \avg{ \vect{F}_n\cdot\bGamma_{m} } &= \frac{\alpha n^2(n+1)}{2n+1} \left( n(n+1){\lambda}-1 \right) \, , \notag \\
 \avg{ \vect{G}_n\cdot\bGamma_{m} } &= \frac{\alpha n(n+1)}{2(2n+1)(2n+3)}  \notag \\
 &\times \big( 12+22n+13n^2+2n^3+2n^2(2n+3){\lambda} \big) \, , \notag \\
 \avg{ \vect{f}_n\cdot\bGamma_{m} } &= -\frac{\alpha n^2(n+1)}{2n+1} \left( 2{\lambda}-1 \right) \, , \notag
\end{align}
where $i\alpha = 2\kS/(3\eta\omega)$.
Combining these with Eqs.~\eqref{Psi_DOT_Psi} through \eqref{Psi_DOT_Gamma}, we get
\begin{align} 
 \label{eq_4}
 &(n+1)\tilde{\alpha}_n+ \tilde{\beta}_n = \alpha \Big[ \left[ (n+1)\alpha_n+\beta_n \right] \\ \nonumber
 &\times \left[ n(n+1){\lambda}-1 \right] - (n+1)(2{\lambda}-1) \rho_n \Big] \, ,  \\ \label{eq_5}
 &\tilde{\alpha}_n +\frac{4(n+1)^2}{(2n+3)n^2} \tilde{\beta}_n = \alpha \Big[ \left( n(n+1){\lambda}-1 \right) \alpha_n \\ \nonumber 
 &+ \frac{12+22n+13n^2+2n^3+2n^2(2n+3){\lambda}}{2n(2n+3)} \beta_n- (2{\lambda}-1)\rho_n \Big] \, .  
\end{align}

Further,  the normal traction jump, given by Eqs.~\eqref{BC:sigma_r_r} and \eqref{jump_u_R_Shear_main}, can be written as 
\begin{equation}
 \infSumOne \left( p_n-p_n^{(i)} \right) \psi_{n-1} = \alpha (2{\lambda}-1) \infSumOne \left[ \rho_{n,r}-(n+1)\rho_n \right] \psi_{n-1} \, , \notag
\end{equation}
leading directly to
\begin{equation}
  p_n-p_n^{(i)} = \alpha (2{\lambda}-1) \left[ \rho_{n,r}-(n+1)\rho_n \right]  \, . \label{eq_6}
\end{equation}

Eqs.~\eqref{eq_4}, \eqref{eq_5} and \eqref{eq_6} together with \eqref{eq_1} through \eqref{eq_3} form a closed system of equations amenable to immediate resolution by the standard substitution method. Finally, we obtain
\begin{equation}
 A_n = \frac{\alpha n}{K} \left( \frac{K_1}{R^{n+1}} - \frac{K_3}{R^{n+3}} \right) \, , \label{A_n_Shearing}
\end{equation}
with the auxiliary functions
\begin{align}
 K_1 &= (2n+3) (n-1) \big[ (4-\alpha)(n^2+4n+3)+3 \notag \\ \nonumber
 &+ \left( 2n^2+(2\alpha+5)n+6\alpha \right)(n+1){\lambda}\big] \, , \\
 K_3 &= (2n+1) (n+1) \big[ (4-\alpha)(n^2+4n+3)+3  \notag \\ \nonumber
 &+\left( 2n^2+(2\alpha+7)n+6\alpha+6 \right)(n+1){\lambda} \big] \, , \\ \nonumber
 K   &= 8{\lambda}\alpha n^5 + 2\left[ (2{\lambda}-1)\alpha^2+30{\lambda}\alpha+16 \right]n^4 \notag \\ \nonumber
  &+ 4\left[ 3(2{\lambda}-1)\alpha^2+43{\lambda}\alpha+48 \right]n^3  \\ 
     &+2 \left[ 11(2{\lambda}-1)\alpha^2+117{\lambda}\alpha+200 \right]n^2 \notag \\ \nonumber
     &+6 \left[ 2(2{\lambda}-1)\alpha^2+(25{\lambda}-2)\alpha+56 \right]n \\ \nonumber
     &+ 18(2{\lambda}-1)\alpha+90 \, .
\end{align}

Further, we express $B_n$ in terms of $A_n$ and $A_{n+2}$ as
\begin{align}  
\label{B_n_Shearing}
  B_n &= -\frac{n+1}{2(n+3)(2n+5)}A_{n+2} \nonumber \\ 
  &+\frac{1}{2G} \Bigg[ \frac{1}{2n+1}\left(G' A_n + \frac{\alpha n G_1}{R^{n+1}}\right) \\ 
  & - \frac{\alpha (n+1) G_3}{(n+3)(2n+5) \left( \alpha n^2+(5\alpha+4)n+4\alpha+10 \right)} \frac{1}{R^{n+3}} \Bigg] \, , \nonumber
\end{align}
with 
\begin{align}
 G   &= {\lambda}\alpha n^3+\left( (6{\lambda}-1)\alpha+4 \right)n^2  \notag \\ \nonumber
 &+\left[ (11{\lambda}-4)\alpha+16 \right]n+3(2{\lambda}-1)\alpha + 15 \, , \\
 G'  &= {\lambda}\alpha n^3 +\left[ (4{\lambda}-1)\alpha+4 \right]n^2 \notag \\ \nonumber
 &+\left[ (5{\lambda}-4)\alpha+8 \right]n+(2{\lambda}-1)\alpha+3 \, , \\
 G_1 &= {\lambda}n^2+n-({\lambda}+1) \, , \notag \\ \nonumber
 G_3 &= {\lambda}\alpha n^5+\left[ (7\lambda+1)\alpha+4{\lambda} \right]n^4 \notag \\ \nonumber
 &+ \left[ 3({\lambda}+4)\alpha+2(17{\lambda}-6) \right]n^3  \\ 
     &+ \left[ 2(52{\lambda}-47)-(71{\lambda}-51)\alpha \right]n^2 \notag \\ \nonumber
     &+ 2\left[ 67{\lambda}-171 - 2(41{\lambda}-22)\alpha \right]n \\ \nonumber 
     &-48(2{\lambda}-1)\alpha + 30(2{\lambda}-11) \, . 
\end{align}
The last coefficient, $C_n$, is found as
\begin{equation}
 C_n = \frac{2}{n+2} A_{n+1} + \frac{2n(n+3) \alpha}{(n+2) \left( \alpha n^2+(3\alpha+4)n+6 \right)} \frac{1}{R^{n+2}} \, . 
 \label{C_n_Shearing}
\end{equation}

In particular, for $\alpha\to\infty$ (obtained either by considering an infinite shearing modulus or a vanishing forcing frequency), we recover the coefficients near a hard-sphere with stick boundary conditions, namely
\begin{align}
 \lim_{\alpha\to\infty} A_n &= \frac{1}{2(n+2)} \bigg( \frac{(2n+3)(n-1)}{R^{n+1}} \notag \\ \nonumber
 &- \frac{(2n+1)(n+1)}{R^{n+3}} \bigg) \, , \\
 \lim_{\alpha\to\infty} B_n &= \frac{n-1}{4(n+2)} \frac{1}{R^{n+1}} + \frac{n+1}{4(n+4)}\frac{1}{R^{n+5}} \notag \\ \nonumber
 & - \frac{n^2+3n-1}{2(n+2)(n+4)}  \frac{1}{R^{n+3}}  \, , \\
 \lim_{\alpha\to\infty} C_n &= \frac{2n+3}{n+3} \left( \frac{1}{R^{n+2}} - \frac{1}{R^{n+4}} \right) \, , \notag
\end{align}
all in agreement with the results of Fuentes {\textit{et al.}} \cite{fuentes89}, and as given in Kim and Karrila \cite[p. 246]{kim13}.


\subsection*{Pure bending}

In complete analogy with the previous section, Eqs.~\eqref{BC:sigma_r_phi} and \eqref{BC:sigma_r_the}  with the right-hand side given by \eqref{jump_u_Phi_Bending_main} and \eqref{jump_u_The_Bending_main}, respectively, can be written as
\begin{equation}
 \infSumOne \tilde{\alpha}_n \bPsi_{n-1}+\infSum \tilde{\beta}_n \bGamma_n = \infSumOne \rho_n \vect{g}_n \, , \label{tangentialTractionJump_Bending}
\end{equation}
where $\vect{g}$ can directly be determined by identification.
Multiplying both members of Eq.~\eqref{tangentialTractionJump_Bending} by $\bPsi_{m-1}$ and by $\bGamma_m$, 
averaging over the surface of the sphere upon making use of the following orthogonality relations 
\begin{align}
 \avg{ \vect{g}_n\cdot\bPsi_{m-1} } &=-\alphaB \frac{(n^2-1)n^2(n+1)(n+2)}{2n+1} \delta_{mn} \, , \notag \\ 
 \avg{ \vect{g}_n\cdot\bGamma_{m} } &=-\alphaB \frac{(n^2-1)n^2(n+2)}{2n+1} \delta_{mn}  \, , \notag
\end{align}
together with Eqs.~\eqref{Psi_DOT_Psi} through \eqref{Psi_DOT_Gamma}, we get
\begin{align}
  \tilde{\alpha}_n+\frac{1}{n+1}\tilde{\beta}_n &= -\alphaB  (n-1)(n+2) \rho_n \, , \label{eq_4_bending} \\
 \tilde{\alpha}_n + \frac{4(n+1)^2}{n^2(2n+3)} \tilde{\beta}_n &= -\alphaB (n-1)(n+2) \rho_n  \, , \label{eq_5_bending}
\end{align}
where $i\alphaB = \kB/(\eta\omega)$.

For the normal traction jump, Eq.~\eqref{BC:sigma_r_r} with \eqref{jump_u_R_Bending_main} can be written as
\begin{equation}
 -\infSumOne \left( p_n-p_n^{(i)} \right) \psi_{n-1} =  \infSumOne \rho_n H_n \, . \notag
\end{equation}
After making use of the orthogonality property 
\begin{equation}
 \avg{ H_n \psi_{m-1} } = - \alphaB \frac{(n^2-1)n^2(n+1)(n+2)}{2n+1} \delta_{mn} \, , \notag
\end{equation}
we obtain
\begin{equation}
 p_n-p_n^{(i)} = -\alphaB (n^2-1)n(n+2) \rho_n \, . \label{eq_6_bending}
\end{equation}

Solving the system of equations formed of Eqs.~\eqref{eq_4_bending}, \eqref{eq_5_bending} and \eqref{eq_6_bending} together with \eqref{eq_1} through \eqref{eq_3}, we obtain the first set of coefficients as
\begin{equation}
 A_n = \frac{\alphaB w}{W} \left[ \frac{(2n+5)(n-1)}{R^{n+1}} - \frac{(2n+1)(n+1)}{R^{n+3}} \right] , \label{A_n_Bending}
\end{equation}
where
\begin{align}
 w &=n^2(n+1)(n+2)(n+3), \notag \\ 
 W &= 30+(12\alphaB +92)n+(94\alphaB+72)n^2 \notag \\ \nonumber
 &+(168\alphaB+16)n^3+ 118\alphaB n^4+36\alphaB n^5+4\alphaB n^6.
\end{align}
For the set $B_n$, we find
\begin{align} \label{B_n_Bending}
 B_n &= -\frac{n+1}{2(n+3)(2n+5)}A_{n+2} + \frac{1}{S} \bigg[ \frac{S' A_n}{2n+1} \\ \nonumber
 &+ \alphaB n(n+3) \left( \frac{(n+1)^2}{2n+5} \frac{1}{R^{n+3}} - \frac{n^2-1}{2n+1} \frac{1}{R^{n+1}} \right) \bigg] \, , 
\end{align}
where we defined
\begin{align}
 S  &= 2 \big[ \alphaB n^4+6\alphaB n^3+(11\alphaB+4)n^2 \notag \\ \nonumber 
 &+2(3\alphaB+8)n+15 \big] \, , \\
 S' &= \frac{S}{2} - 8n - 12 \, . \notag
\end{align}
Finally, the last set is simply given by
\begin{equation}
 C_n = \frac{2 A_{n+1}}{n+2} \, . \label{C_n_Bending}
\end{equation}

The same resolution procedure can be applied to the evaluation of the series coefficients when the membrane is endowed simultaneously with both shearing and bending resistances.
Analytical expressions can be derived by computer algebra software but they are not listed here due to their complexity and lengthiness.
It is worth to mention that a coupling between shearing and bending exists, i.e. in the same way as observed in part~I \cite{daddi17b} and for two parallel planar membranes \cite{daddi16b} but in contrast to what has been observed for a single membrane \cite{daddi16, daddi17}.

%

\end{document}